\documentclass[12pt]{article}

\usepackage{amssymb,amsfonts,amsthm}
\usepackage{amsmath}
\usepackage{bm}             
\usepackage[mathscr]{eucal}
\usepackage{cancel}
\usepackage{wasysym}

\def\be{\begin{equation}}
\def\ee{\end{equation}}
\def\bea{\begin{eqnarray}}
\def\eea{\end{eqnarray}}

\def\bna{\mbox{\boldmath $\nabla$}}

\def\bdelta{\mbox{\boldmath $\delta$}}

\def\la{\langle}
\def\ra{\rangle}
\def\hsp5{\hspace{5mm}}

\theoremstyle{remark}

\newcommand{\sfrac}[2]{{\textstyle{#1\over#2}}}

\title{\sc Dynamics of cosmological perturbations at first and second order}

\begin{document}

\author{ \\
{\Large\sc Claes Uggla}\thanks{Electronic address:
{\tt claes.uggla@kau.se}} \\[1ex]
Department of Physics, \\
Karlstad University, S-651 88 Karlstad, Sweden
\and \\
{\Large\sc John Wainwright}\thanks{Electronic address:
{\tt jwainwri@uwaterloo.ca}} \\[1ex]
Department of Applied Mathematics, \\
University of Waterloo,Waterloo, ON, N2L 3G1, Canada \\[2ex] }

\date{}
\maketitle

\begin{abstract}

\end{abstract}

In this paper we give five gauge-invariant systems of governing
equations for first and second order scalar perturbations of flat
Friedmann-Lema\^{i}tre universes that are minimal
in the sense that they contain no redundant equations or variables.
We  normalize the variables so that they are dimensionless, which
leads to systems of equations  that
are simple and ready-to-use. We compare the properties and
utility of the different systems. For example, they
serve as a starting point for finding explicit solutions for two
benchmark problems in cosmological perturbation theory at second
order: adiabatic perturbations in the super-horizon regime
(the long wavelength limit) and perturbations of $\Lambda$CDM universes.
However, our framework has much wider applicability and serves as a
reference for future work in the field.

\section{Introduction}

Perturbations of Friedmann-Lema\^{i}tre (FL) cosmologies play an
essential role in confronting theoretical models with observations
of the anisotropy of the cosmic microwave background (CMB) and the
inhomogeneity of the large scale structure (LSS) of the Universe.
Initially linear perturbations were adequate but now the increasing
accuracy of the observations necessitates the use of second order
(nonlinear) perturbations to analyze, for example, the presence of
non-Gaussianity in the CMB and the LSS.\footnote
{See, for example, Bartolo \emph{et al} (2010)~\cite{baretal10} and
Tram \emph{et al} (2016)~\cite{traetal16}.}

In this paper we consider first and second order scalar
perturbations of FL universes subject to the following assumptions:
\begin{itemize}
\item[i)] the spatial background is flat;
\item[ii)] the stress-energy tensor can be written in the form
$T^a\!_b = \left(\rho + p\right)\!u^a u_b + p\delta^a\!_b$,
thereby describing perfect fluids and scalar fields;
\item[iii)] the linear perturbation is purely scalar.
\end{itemize}
The dynamics of perturbations of FL universes are governed by
the perturbed Einstein equations and the perturbed matter equations.
For \emph{scalar perturbations} the perturbed Einstein equations give
four equations (linear combinations of the components of the perturbed
Einstein tensor) which include evolution equations for the metric
perturbations. The perturbed conservation equations provide evolution
equations for two primary matter perturbations, the density
perturbation and the scalar velocity perturbation. Only four of
these six equations are needed to fully describe the perturbations, but
in order to obtain a well-defined system the gauge freedom has to be
eliminated by fixing the gauge.

Since  2004 much work  aimed at confronting theoretical models with
observations has been done using second order perturbations.
In one respect second order perturbations are analogous to first order
perturbations: the \emph{leading order terms} in the equations have
exactly the same form. The  greater complexity at second order arises
from the fact that each equation is augmented by so-called
\emph{source terms} that depend quadratically on the first order
perturbations. There are various ways of formulating the governing
equations for second order perturbations, depending on the choice of
variables and gauge. These choices are influenced by various factors
such as the problem to be investigated, for example, long wavelength
perturbations or perturbations of the $\Lambda CDM$ universe, or in
the case of numerical work, by the availability of numerical packages.
A number of detailed formulations of the governing equations have been given,\footnote
{See for example, Noh and Hwang (2004)~\cite{nohhwa04}, and
Nakamura (2007)~\cite{nak07}.}
but mainly due to
the complexity of the source terms no standard systems have emerged:
it is as though the necessary technical infrastructure for analyzing
second order perturbations has not been sufficiently well developed.

With this as motivation, our goal in this paper is to present five
systems of equations that are suitable for analyzing the dynamics
of both first and second order scalar perturbations of FL universes.
To accomplish this we begin by imposing the so-called C-gauge of Hwang and
Noh~\cite{nohhwa04} up to second order which fixes the spatial
gauge, but we initially keep an
arbitrary temporal gauge. Within this framework we construct
a set of leading order and
quadratic source terms for the perturbed Einstein field equations
(a set of four scalar equations) and the perturbed energy-momentum conservation
equations (a set of two scalar equations).
Finally we construct five specific
systems of gauge invariant equations by
also fixing the temporal gauge. Firstly, specializing the
perturbed Einstein field equations to the Poisson (longitudinal,
zero shear) gauge and the uniform (flat) curvature gauge yields
two systems of governing equations. Secondly, specializing three of the perturbed
Einstein field equations together with the perturbed momentum
conservation equation to the Poisson gauge and the total matter gauge results in
two more systems. Finally we create a fifth system
by using the perturbed energy-momentum equations to describe the
evolution of the density perturbation in the total matter gauge,
and the velocity perturbation in the Poisson gauge, with two
of the perturbed Einstein equations acting as constraints to
determine the metric perturbations.
We regard these five systems of equations
as \emph{ready-to-use}  since they are gauge invariant,
contain no redundant equations or variables,
and do not require that any further simplifications be made before use.

It is important to note that in general the above systems of equations
are not closed (not fully determined) since the non-adiabatic pressure
perturbation has to be specified. However, for a barotropic perfect
fluid and a minimally coupled scalar field, the systems are fully determined,
once an equation of state and a scalar field potential, respectively, has
been given. Moreover, we present the systems in a manner that makes it possible
to apply them to more general matter models
such as multi-fluids and multiple scalar fields.

The present paper is the second of four related papers by the authors.
The first paper~\cite{uggwai19a}, hereafter referred to as UW1,
gives a unified and simplified formulation of
gauge change formulas at second order, while the third paper~\cite{uggwai19b},
called UW3, uses the present
paper, in conjunction with UW1, to give new conserved
quantities and derive the general explicit solution
at second order for adiabatic perturbations in the long wavelength limit,
results that are subsequently adapted to inflationary
universes with a single scalar field in~\cite{uggwai19c}, which we refer to as UW4.

The outline of the paper is as follows. In section~\ref{variables} we
introduce the metric and matter perturbation variables. In
section~\ref{pert.einst} we present leading order and quadratic source terms
for the perturbed Einstein field equations, and in section~\ref{pert_cons}
we present the leading order and quadratic source terms for the perturbed
conservation equations. In both cases the details of the source terms are
deferred to an appendix. The central goal of the paper is reached in
section~\ref{gov.eq} where we derive the five ready-to-use systems of
governing equations. Finally in section~\ref{discussion} we comment on
specific applications of the five systems and on their relative merits.

\section{Metric and matter perturbation variables~\label{variables}}

\subsection{Background geometrical and matter scalars}

The background Robertson-Walker (RW) metric has the form
\begin{equation}  \label{RW}
ds^2 = a^2\left(- d\eta^2 + \gamma_{ij} dx^i dx^j \right),
\end{equation}
where $a$ is the background scale factor, $\eta$ is conformal time
and $\gamma_{ij}$ is the flat spatial 3-metric. The evolution of the background
geometry is governed by the scalars
\begin{subequations}  \label{background.scalars}
\begin{equation}\label{Hq}
{\cal H} = \frac{a'}{a}, \qquad q = -\frac{{\cal H}'}{{\cal H}^2},
\end{equation}
where $'$ denotes differentiation with respect to $\eta$, and
${\cal H} = aH$, with $H$ being the background
Hubble parameter and $q$ the background deceleration parameter.
We associate the following scalars with the background stress-energy tensor:
\begin{equation} \label{RW_matter}
w = \frac{p_0}{{\rho}_0}, \qquad c_s^2 = \frac{p_0'}{{\rho}_0'},
\end{equation}
where $\rho_0$ and $p_0$ are the background energy density and pressure, respectively.
The density parameter is defined as usual by
\begin{equation}
\Omega = \frac{\rho_0}{3H^2},
\end{equation}
\end{subequations}
where we have set $c=1$ and $8\pi G=1$, where
$c$ is the speed of light and $G$ the gravitational constant.

The Einstein equations for a spatially flat background can be written as
\begin{equation}
3{\cal H}^2 = a^2\rho_0, \qquad 2(-{\cal H}'+{\cal H}^2)=a^2(\rho_0 + p_0),
\end{equation}
or equivalently, using equations~\eqref{background.scalars}, in the following form:
\begin{equation}\label{EFE_scale.invariant}
\Omega=1,  \qquad 2(1+q)= 3(1+w).
\end{equation}
One can use the second equation
to switch between $1+q$ and $1+w$ and in what follows we will use
either expression depending on the context.

As regards dimensions, we make the choice that the scale
factor $a$ is dimensionless, which implies via~\eqref{RW} that the conformal time
$\eta$ has dimensions of \emph{length}.
It follows that $H$ and $\cal H$ have dimension $(length)^{-1}$
and $\rho_0$ and $p_0$ have dimension $(length)^{-2}$ while
$q,w,c_s^2$ and $\Omega$ are dimensionless.

We now introduce the background
\emph{$e$-fold time variable} $N$, defined by
\begin{equation}
N = \ln (a/a_0).
\end{equation}
This variable describes the number of
background $e$-foldings with respect to some reference epoch $a=a_0$.
Although conformal time $\eta$ is arguably the most commonly used
background time variable in cosmological perturbation theory, the
$e$-fold time variable $N$ is used in inflationary cosmology
and also when doing numerical simulations.\footnote{See,
for example, Huston and Malik (2009)~\cite{husmal09}.}
In this paper we will primarily use $e$-fold time $N$ but on occasion we
will make the transition to conformal time $\eta$. In
changing time variables, note that
\begin{equation}
\partial_{\eta}= {\cal H}\partial_{N}, \quad
\partial_{\eta}^2= {\cal H}^2(\partial_{N}^2-q\partial_{N}),
\end{equation}
and that the deceleration parameter $q$ can be written in either of the following  forms:
\begin{equation} \label{q.alt}
q= -\frac{\partial_N{\cal H}}{\cal H} = -\frac{\partial_N H}{H} -1.
\end{equation}

In order to write simple expressions for the perturbed Einstein
tensor it is helpful to introduce an additional background geometrical scalar
${\cal C}^2$, which is defined in terms of the background Einstein tensor according to
\begin{equation} \label{cal.C}
{\cal C}^2 = -\sfrac13 \partial_N {}^{(0)}\!G^i\!_i / \partial_N{}^{(0)}\!G^\eta\!_\eta.
\end{equation}
On noting that ${}^{(0)}\!T^i\!_i=3p_0,\,{}^{(0)}\!T^\eta\!_\eta=-\rho_0$
it follows from~\eqref{RW_matter}  that ${\cal C}^2$ is the
geometrical analogue of $c_s^2$ and that
the Einstein equations in the background imply
\begin{equation}
{\cal C}^2 = c_s^2.
\end{equation}
For future use we note that the definition~\eqref{cal.C} leads to
the following derivative:\footnote
{Write the background Einstein tensor in the form
${}^{(0)}\!G^\eta\!_\eta=-3H^2,\, {}^{(0)}\!G^i\!_i=3H^2 (1-2q).$}
\begin{equation} \label{deriv_q}
\partial_N q= - (1+q)(1 + 3{\cal C}^2 - 2q).
\end{equation}
%

\subsection{Metric perturbation variables}

To perturb a flat RW background geometry we write the metric in the form
\begin{equation}\label{metric_exp}
ds^2 = a^2\left(-(1+2\phi) d\eta^2 +  f_{\eta i}\,d\eta dx^i + f_{ij} dx^i dx^j \right),
\end{equation}
where we assume that the metric components can be expanded as a Taylor series in
a perturbation parameter $\epsilon$, e.g.,
\begin{equation}
\phi = \epsilon\,{}^{(1)}\!\phi + \sfrac12 \epsilon^2\,{}^{(2)}\!\phi + \dots\, .
\end{equation}
We also assume that the metric can be decomposed into scalar, vector, and tensor perturbations according to
\begin{subequations}\label{metric}
\begin{align}
f_{\eta i} &=  {\bf D}_i B + B_i, \label{metric.B} \\
f_{ij} &= (1-2\psi)\gamma_{ij}+ 2{\bf D}_i {\bf D}_j C + 2{\bf D}_{(i }C_{j)} + 2C_{ij},\label{metericfij}
\end{align}
\end{subequations}
where ${\bf D}^iB_i = 0$, ${\bf D}^iC_i = 0$, $C^i\!_i=0$, ${\bf D}^iC_{ij} = 0$,
which ensures that $\psi, B$ and $C$ describe scalar perturbations. Here
${\bf D}_i$ is the spatial covariant derivative
corresponding to the flat metric $\gamma_{ij}$.
Use of Cartesian background coordinates yields
$\gamma_{ij} = \delta_{ij}$ and ${\bf D}_i = \partial/\partial x^i$.
As regards dimensions, since we have made the choice that the scale factor $a$ is dimensionless,
it follows that the coordinates $\eta$ and $x^i$ have dimensions of \emph{length}
since $ds^2$ has dimension $\text{\emph{length}}^2$.
Therefore, due to the structure of equations~\eqref{metric_exp} and~\eqref{metric},
$\phi$ and $\psi$ are dimensionless while $B$ has
dimension \emph{length}, since ${\bf D}_i $ has dimension $(\emph{length})^{-1}$.


From now on we completely fix the spatial gauge freedom
by setting the metric functions $C$ and
$C_i$ in~\eqref{metric} to be zero order by order,\footnote{This excludes the synchronous gauge
(for a recent work using the synchronous gauge, see e.g.~\cite{grebru17}), but apart from
this gauge most commonly used gauges are included in this class. }
which up to second order gives
\begin{equation}\label{Ccond}
{}^{(r)}\!C = 0, \qquad {}^{(r)}\!C_i = 0,\qquad r=1,2,
\end{equation}
where the transformation laws for $C$ and $C_i$ were given in equations (B10e)
and (B10f) in~\cite{uggwai14a}.
Furthermore, in this paper we are restricting our considerations to perturbations
that are \emph{purely scalar at linear order}, and hence the metric perturbations
$B_i$ and $C_{ij}$ satisfy\footnote{This assumption is often made,
but see for example Carrilho and Malk (2016)~\cite{carmal15}.}
\begin{equation}
{}^{(1)}\!B_i = 0, \qquad {}^{(1)}\!C_{ij} = 0.
\end{equation}
Since purely scalar perturbations at linear order will generate vector and tensor perturbations
at second order it follows that these perturbations will have
${}^{(2)}\!B_i \neq 0$, ${}^{(2)}\!C_{ij} \neq 0$. However, our
interest in this paper is in discussing the scalar perturbations at first and second order.
Thus the metric perturbations we consider are given by
\begin{equation} \label{metric_var}
{}^{(r)}\!f  = ({}^{(r)}\!\phi,\, {\cal H}{}^{(r)}\!B, {}^{(r)}\!\psi), \qquad r=1,2.,
\end{equation}
where we have scaled ${}^{(r)}\!B$ with a factor of $\cal H$. We
introduced this scaling in our earlier paper UW1~\cite{uggwai19a},
 motivated by the transformation properties of the metric perturbations
under a change of gauge (see equations (23) and (24) in that paper),
and by the fact that  ${\cal H}{}^{(r)}\!B$ is dimensionless, as we have confirmed here.
By inspection it follows from~\eqref{metric_exp} and~\eqref{metericfij} that $\phi$
and $\psi$ are dimensionless.

For future reference when
scaling variables with ${\cal H}$ we use
\begin{equation} \label{scale.H}
{\cal H}\partial_{N}f= (\partial_{N}+q)({\cal H}f),
\end{equation}
as follows from~\eqref{q.alt}.

\subsection{Matter perturbation variables}

We consider a stress-energy tensor which can be written on the form
\begin{equation}\label{pf}
T^a\!_b = \left(\rho + p\right)\!u^a u_b + p\delta^a\!_b,
\end{equation}
which describes both perfect fluid models and models with a minimally coupled
scalar field. In addition we assume that it can be expanded in a Taylor series in $\epsilon$,
\emph{e.g.}
\begin{equation}
\rho(\epsilon) = \rho_0 + \epsilon\,{}^{(1)}\!\rho +
\sfrac12\epsilon^2\,{}^{(2)}\!\rho + \dots,  \label{rho_exp}
\end{equation}
for the energy density and similarly for the pressure $p(\epsilon)$. We then
normalize the perturbations of $\rho$ and $p$ with
$\rho_0 +p_0$ and define
\begin{equation} \label{rho,p_normalized}
{}^{(r)}\!{\bdelta} = \frac{{}^{(r)}\!\rho}{\rho_0 + p_0}, \qquad
{}^{(r)}\!{P} = \frac{{}^{(r)}\!p}{\rho_0 + p_0},
\end{equation}
which are dimensionless.
%
%

To define the scalar velocity perturbations we find it convenient to
work with the \emph{covariant} 4-velocity $u_b$, which we normalize
with a conformal factor $a$ according to $u_b=a V_b$, in analogy
with the conformal factor $a^2$ in the metric~\eqref{metric_exp}.
We then expand and decompose the
spatial components of $V_b$ according to
\begin{subequations} \label{vel_exp}
\begin{align}
V_i &= \epsilon\,{}^{(1)}\!{V}_i +
\sfrac12 \epsilon^2\,{}^{(2)}\!{V}_i + \dots, \label{vel_exp1} \\
{}^{(r)}\!{V}_i &= {\bf D}_j {}^{(r)}\!{V} + {}^{(r)}\!{\tilde V}_i,\qquad r = 1, 2,\dots ,
\end{align}
\end{subequations}
with ${\bf D}^i{}^{(r)}\!{\tilde V}_i=0$, so that ${}^{(r)}\!{V}$
represents the scalar perturbations. Since we are focussing on scalar
perturbations in this paper we
set the first order vector term to zero.
Since $V_i$ is dimensionless and the $x^i$ have dimension
\emph{length} it follows from~\eqref{vel_exp}
that ${}^{(r)}\!V$ has dimension \emph{length}.
As in the case of ${}^{(r)}\!B$
we normalize ${}^{(r)}\!{V}$ with ${\cal H}$ and thereby consider
${\cal H}{}^{(r)}\!{V}$.

Next we introduce the non-adiabatic pressure perturbations,
which we denote by ${}^{(r)}\!{\Gamma}, \,r=1,2$. Following
Bartolo \emph{et al} (2004)~\cite{baretal04a}, but using the
normalized pressure perturbation~\eqref{rho,p_normalized},
we define\footnote{See in particular~\cite{baretal04a}
equations (136), (137) and (145) for a general discussion and
the desired expressions. The Bartolo expression $\delta P_{nad}$
is related to our $\Gamma$ by $\delta P_{nad}=(\rho_0+p_0)\Gamma.$
Previously, Kodama and Sasaki (1984)~\cite{kodsas84} used the symbol
$\Gamma$ in this context: their expression is related to ours according to
$w\Gamma_{KS}=(1+w)\Gamma.$ See equation (II.3.38).}
\begin{equation}
{}^{(r)}\!{\Gamma} = {}^{(r)}\!{P}_{\rho}, \qquad r=1,2,  \label{gamma_def}
\end{equation}
\emph{i.e. the non-adiabatic pressure perturbations equal
the pressure perturbations in the uniform density gauge} (defined by
${}^{(r)}\!{\bdelta}=0,\, r=1,2$). This definition ensures
that the ${}^{(r)}\!{\Gamma}, \,r=1,2$, are dimensionless gauge invariants and that  if
$p=p(\rho)$ then ${}^{(r)}\!{\Gamma} =0$.
Since we will need to express ${}^{(r)}\!{\Gamma}$, $r=1,2$, in terms
of other spatially fixed gauges we use change of gauge formula
to express ${}^{(r)}\!{P}_{\rho}$, $r=1,2$, in terms of
the normalized pressure and density perturbations
in a temporally arbitrary but spatially fixed gauge.\footnote{See~Malik and Wands
(2004)~\cite{malwan04}, equations (4.19) and (4.20),
and~\cite{baretal04a}, equations (137) and (145)}
On introducing the scaled density perturbation~\eqref{rho,p_normalized}
and the $e$-fold time variable $N$ we obtain the following expressions:
\begin{subequations} \label{gamma_general}
\begin{align}
{}^{(1)}\!{\Gamma} &= {}^{(1)}\!{P} - c_s^2\, {}^{(1)}\!{\bdelta},\label{gamma_1} \\
{}^{(2)}\!{\Gamma} &= {}^{(2)}\!{P} - c_s^2\,{}^{(2)}\!{\bdelta} +
\sfrac13 (\partial_N c_s^2){}^{(1)}\!{\bdelta}^2 +
\sfrac23 {}^{(1)}\!{\bdelta}\left[\partial_N -
3(1 + c_s^2)\right]\!{}^{(1)}\!{\Gamma}. \label{gamma_2}
\end{align}
\end{subequations}
%

In what follows we will replace the pressure perturbations ${}^{(r)}\!{P}$
with the gauge invariants ${}^{(r)}\!{\Gamma}$, $r=1,2$,
which means that the basic matter
perturbations that we use are the dimensionless quantities
\begin{equation}\label{matter_var}
{}^{(r)}\!M  = ({}^{(r)}\!{\bdelta}, {}^{(r)}\!\Gamma, {\cal H}{}^{(r)}\!V),\qquad r=1,2.
\end{equation}
%


\section{The perturbed Einstein equations in a general temporal gauge\label{pert.einst}}

\subsection{Perturbed Einstein and stress energy tensors~\label{GandT_pert} }

We assume that the Einstein tensor for the
metric~\eqref{metric_exp} has a Taylor expansion of the form
\begin{equation}
G^a\!_b(\epsilon) = {}^{(0)}\!G^a\!_b + \epsilon\, {}^{(1)}\!G^a\!_b +
\sfrac12 \epsilon^2\,{}^{(2)}\!G^a\!_b + \dots\, .  \label{Gab_exp}
\end{equation}
The first and second order perturbations of the Einstein
tensor have the following general structure:
\begin{subequations}  \label{pert_Gab}
\begin{align}
H^{-2}\,{}^{(1)}\!G^a\!_b &= {\sf G}^a\!_b({}^{(1)}\!f),   \\
H^{-2}\,{}^{(2)}\!G^a\!_b &= {\sf G}^a\!_b({}^{(2)}\!f) + {\mathbb G}^a\!_b({}^{(1)}\!f),
\end{align}
\end{subequations}
where we have normalized with the background quantity
$H^{-2}$ to ensure that ${\sf G}^a\!_b(f)$ and
${\mathbb G}^a\!_b({}^{(1)}\!f)$ are dimensionless.
Observe that the first and second order perturbations
have a common \emph{leading order term} of the form ${\sf G}^a\!_b(f)$, where $f$ denotes
${}^{(1)}\!f \equiv ({}^{(1)}\!\phi,{\cal H}{}^{(1)}\!B, {}^{(1)}\!\psi)$ or
${}^{(2)}\!f \equiv ({}^{(2)}\!\phi,{\cal H}{}^{(2)}\!B, {}^{(2)}\!\psi)$,
while $H^{-2}\,{}^{(2)}\!G^a\!_b$
also has a \emph{source term} ${\mathbb G}^a\!_b({}^{(1)}\!f)$ which depends quadratically
on ${}^{(1)}\!f $.
We use the fonts ${\cal G}$, ${\sf G}$, ${\mathbb G}$ as notational conventions
for background (0th order), first/leading order,
and second order source terms, respectively, although we
will deviate from these conventions when they clash
with notation that has become fairly standard in the literature.

The perturbations of the stress-energy tensor have a similar general structure:
\begin{subequations} \label{pert_Tab}
\begin{align}
(\rho_0 + p_0)^{-1}\,{}^{(1)}\!T^a\!_b &= {\sf T}^a\!_b({}^{(1)}\!M),  \\
(\rho_0 + p_0)^{-1}\,{}^{(2)}\!T^a\!_b &= {\sf T}^a\!_b({}^{(2)}\!M) +
{\mathbb T}^a\!_b({}^{(1)}\!f, {}^{(1)}\!M),
\end{align}
\end{subequations}
where ${}^{(r)}\!M  = ({}^{(r)}\!{\bdelta}, {}^{(r)}\!\Gamma, {\cal H}{}^{(r)}\!V)$, $r=1,2$,
are the matter perturbation variables. We have chosen the normalization
factor $( \rho_0 +p_0)^{-1}$ to be compatible with~\eqref{rho,p_normalized},
resulting in dimensionless variables.

We express the perturbed Einstein equations ${}^{(r)}\!G^a\!_b={}^{(r)}\!T^a\!_b$, $r=1,2$,
in leading order terms and source terms using equations~\eqref{pert_Gab}
and~\eqref{pert_Tab}. Since $(\rho_0 + p_0)/H^2 = 3(1+w)\Omega = 3(1+w)$ the first
and second order perturbed Einstein equations take the form:
\begin{subequations} \label{einst_pert_gen}
\begin{align}
{\sf G}^a\!_b({}^{(1)}\!f) &= 3(1+w)\,{\sf T}^a\!_b({}^{(1)}\!M),  \label{epg1}  \\
{\sf G}^a\!_b({}^{(2)}\!f) + {\mathbb G}^a\!_b({}^{(1)}\!f) &=
3(1+w)\left({\sf T}^a\!_b({}^{(2)}\!M) +
{\mathbb T}^a\!_b({}^{(1)}\!f, {}^{(1)}\!M)\right). \label{epg2}
\end{align}
\end{subequations}
%

\subsection{The scalar mode}

The scalar mode of the leading order tensor
${\sf G}^a\!_b(f)$ in~\eqref{pert_Gab}  is described
by the following linear combinations of ${\sf G}^a\!_b$:\footnote{The
components ${\sf G}_{ij}$ contain a vector mode and a tensor
mode in addition to the scalar mode. The operator ${\cal S}^{ij}$ is a concise
way of extracting the scalar mode. Similarly the operator   ${\cal S}^i$
extracts the scalar mode from ${\sf G}^\eta\!_i$.}
\begin{equation} \label{scalar_G}
{\sf G}^i\!_i, \quad {\cal H}^2{\cal S}^{ij}{\sf G}_{ij}, \quad
{\cal H}{\cal S}^i{\sf G}^\eta\!_i, \quad {\sf G}^\eta\!_\eta,
\end{equation}
and similarly for the leading order tensor ${\sf T}^a\!_b(M)$ in~\eqref{pert_Tab}
and the source terms ${\mathbb G}^a\!_b$ and ${\mathbb T}^a\!_b$.
Here the \emph{scalar mode extraction operators} ${\cal S}^i$
and ${\cal S}^{ij}$, see Uggla and Wainwright
(2013)~\cite{uggwai13a}, are defined as follows:
\begin{subequations}\label{modeextractop}
\begin{align}
{\cal S}^{i} &= {\bf D}^{-2}{\bf D}^i ,\\
{\cal S}^{ij} &= \sfrac32({\bf D}^{-2})^2{\bf D}^{ij},
\end{align}
\end{subequations}
where ${\bf D}_{ij} = {\bf D}_{(i}{\bf D}_{j)} - \sfrac13 \gamma_{ij}{\bf D}^2$,
${\bf D}^2$ is the spatial Laplacian and ${\bf D}^{-2}$ is the inverse Laplacian.
To ensure that the expressions~\eqref{scalar_G} are dimensionless
we have scaled ${\cal S}^i$ and ${\cal S}^{ij}$ appropriately with ${\cal H}$
(see also~\eqref{GandT_lincombo} below and appendix~\ref{source}).

We have found that significant simplifications occur in the
perturbed Einstein equations if one replaces ${\sf G}^i\!_i$ and ${\sf T}^i\!_i$
by the following combinations :
\begin{equation}
\sfrac13{\sf G}^i\!_i + {\cal C}^2 {\sf G}^\eta\!_\eta, \qquad
\sfrac13{\sf T}^i\!_i + c_s^2\, {\sf T}^\eta\!_\eta,
\end{equation}
where ${\cal C}^2$ and $c_s^2$ are the background scalars
defined by equations~\eqref{RW_matter} and~\eqref{cal.C},
with ${\cal C}^2=c_s^2$ when the background
Einstein equations are satisfied. The motivation for this choice
is clear in the case of the stress-energy tensor, since it follows
that $\sfrac13{\sf T}^i\!_i(M) + c_s^2\, {\sf T}^\eta\!_\eta(M)=\Gamma,$
the non-adiabatic pressure perturbation. Consistency then requires that
we use $\sfrac13{\sf G}^i\!_i + {\cal C}^2 {\sf G}^\eta\!_\eta$ for
the Einstein tensor.

With the above as motivation we now define
the following linear combinations of the components of the leading order tensors
${\sf G}^a\!_b(f)$ and ${\sf T}^a\!_b(M)$ for scalar perturbations:
\begin{subequations} \label{GandT_lincombo}
\begin{xalignat}{2}
{\sf G}^{\Gamma}(f) &:= \sfrac13{\sf G}^i\!_i + {\cal C}^2 {\sf G}^\eta\!_\eta, &\quad
{\sf T}^{\Gamma}(M) &:= \sfrac13{\sf T}^i\!_i + c_s^2\, {\sf T}^\eta\!_\eta = \Gamma, \\
{\sf G}^{\pi}(f) &:= {\cal H}^2{\cal S}^{ij}{\sf G}_{ij}, &\quad
{\sf T}^{\pi}(M) &:= {\cal H}^2{\cal S}^{ij}{\sf T}_{ij} = 0,\\
{\sf G}^q(f) &:= {\cal H}{\cal S}^i{\sf G}^\eta\!_i, &\quad
{\sf T}^q(M) &:= {\cal H}{\cal S}^i{\sf T}^\eta\!_i = {\cal H}V,\\
{\sf G}^{\rho}(f) &:= -{\sf G}^\eta\!_\eta, &\quad
{\sf T}^{\rho}(M) &:= -{\sf T}^\eta\!_\eta = \bdelta,
\end{xalignat}
\end{subequations}
where $f={}^{(1)}\!f$ or ${}^{(2)}\!f$ and $M={}^{(1)}\!M$ or ${}^{(2)}\!M$,
as given by~\eqref{metric_var}  and~\eqref{matter_var},
respectively. We will use the same linear combinations for the
source terms ${\mathbb G}^a\!_b({}^{(1)}\!f)$ and ${\mathbb T}^a\!_b({}^{(1)}\!f,{}^{(1)}\!M)$,
which are given by the above equations with ${\sf G}$ and ${\sf T}$
replaced by ${\mathbb G}$ and ${\mathbb T}$, respectively.

\subsection{The leading order Einstein tensor terms}

The expressions for the leading order Einstein terms
in~\eqref{GandT_lincombo} can be obtained
by specializing equations (19) in
Uggla and Wainwright (2013)~\cite{uggwai13a},
which yields:\footnote{The notation
in~\cite{uggwai13a} is related to the notation in the present paper as follows:
${\sf G}={\cal H}^2{\sf G}^{\Gamma}$, ${\cal S}^{ij}{\hat{\sf G}}_{ij}={\sf G}^{\pi}$,
${\cal S}^i{\sf G}_i = {\cal H}^2({\sf G}^{\rho}-3{\sf G}^q)$,
and ${\cal S}^i{\sf G}^0\!_i =  {\cal H}{\sf G}^q$. The differential operators have been scaled and relabelled as
${\cal L}_A\equiv {\cal H}{\cal L}_1$, ${\cal L}_B \equiv {\cal H}{\cal L}_2$
and ${\cal L} \equiv {\cal H}^2{\cal L}_B$.
Specialize $f_{ab}$ according to $f_{00}=-2\phi$, $f_{0i}= {\bf D}_i B$, $f_{ij}=-2\psi \gamma_{ij}$, and set
$K=0$, since we are considering a flat background. \label{link.2013}}
\begin{subequations} \label{lead_Gab}
\begin{align}
\begin{split}
\sfrac12{\mathsf G}^{\Gamma}(f) + \sfrac13{\cal H}^{-2}{\bf D}^2{\mathsf G}^{\pi}(f) &= \\
({\cal L}_\mathrm{B} - {\cal C}^2{\cal H}^{-2}{\bf D}^2)\psi  & +
{\cal L}_1 (\phi -\psi) + {\cal C}^2{\cal H}^{-2}{\bf D}^2({\cal H}{B}) , \end{split} \label{lG.1}  \\
{\mathsf G}^{\pi}(f) &= - ({\cal L}_2 + q)({\cal H}{B}) - \phi + \psi, \label{lG.2} \\
{\mathsf G}^q(f) &= -2\left(\partial_N\psi + \phi\right),  \label{lG.3}\\
{\mathsf G}^{\rho}(f) - 3{\mathsf G}^q(f) &=
 2{\cal H}^{-2}{\bf D}^2 \left(\psi - {\cal H}{B}\right),  \label{lG.4}
\end{align}
\end{subequations}
where $f = {}^{(1)}\!f$ or $f = {}^{(2)}\!f$.
The temporal differential operators ${\cal L}_1$, ${\cal L}_2$,
which are first order in time, are defined by
\begin{subequations}\label{L_1,2}
\begin{xalignat}{2}
{\cal L}_1 &= \partial_N  + 1 + 3{\cal C}^2 - 2q, &\qquad  {\cal L}_1f &= (1+q)\partial_N\left((1+q)^{-1} f\right), \label{L1}  \\
{\cal L}_2 &= \partial_N  + 2, &\qquad  {\cal L}_2 f &= a^{-2}\partial_N(a^2 f), \label{L2}
\end{xalignat}
\end{subequations}
where the compact product form expressions are derived using
$\partial_N a=a$ and the derivative~\eqref{deriv_q} of $q$.
The Bardeen operator ${\cal L}_\mathrm{B}$,
which is of second order, is defined as:
\begin{equation}\label{LB}
{\cal L}_\mathrm{B}(f) = {\cal L}_1({\cal H}{\cal L}_2({\cal H}^{-1}f)),
\end{equation}
where $f$ is an arbitrary function. Expanding the product
form~\eqref{LB} using~\eqref{L_1,2}, and using the
definition~\eqref{Hq} of $q$ and the derivative~\eqref{deriv_q} of $q$, leads to
\begin{equation} \label{LB_expand}
{\cal L}_\mathrm{B} = \partial_N^2 + (3(1 + {\cal C}^2) - q)\partial_N +
1 + 3{\cal C}^2 - 2q.
\end{equation}
The operators ${\cal L}_B,\,{\cal L}_1$ and ${\cal L}_2$
play a central role in determining the evolution of scalar perturbations
at first and second order. Note that they are purely kinematical in nature,
and therefore relevant for any metric theory that involves the Einstein tensor.
The operator ${\cal L}_B$ is associated with
the Poisson gauge, and gained prominence through the seminal paper
of Bardeen (1980)~\cite{bar80}, while the
operators ${\cal L}_1$ and ${\cal L}_2$
are associated with the uniform curvature gauge and
the work of Kodama and Sasaki~\cite{kodsas84} (see equations (4.6a,b)),
but have been less used.
The scalars $q$ and ${\cal C}^2$ are determined by
the background Einstein equations once the stress-energy
tensor has been specified, for example a perfect fluid
with barotropic equation of state, pressure-free matter (cold dark matter (CDM)) with
a cosmological constant, or a minimally coupled scalar field. In this way
the Bardeen operator has appeared in the literature in a variety of different forms,
usually using conformal time as the time variable:\footnote{The third
term on the right side appears in different forms, for example
${\cal H}^2\!\left(1 + 3{\cal C}^2 - 2q\right)=2{\cal H}' +
{\cal H}^2(1 + 3c_s^2)=3{\cal H}^2(c_s^2-w),$
for a perfect fluid universe, (\emph{e.g.} Mukhanov \emph{et al} (1992)~\cite{muketal92},
equation (5.22), Nakamura (2007)~\cite{nak07}, equation (6.65),
Malik and Wands (2009)~\cite{malwan09}, equation (8.31))}
\begin{equation} \label{LB_eta}
{\cal L}_\mathrm{B} ={\cal H}^{-2}\left( \partial_{\eta}^2 +
3(1 + {\cal C}^2){\cal H} \partial_{\eta} +{\cal H}^2(1 + 3{\cal C}^2 - 2q)\right).
\end{equation}
\subsection{The perturbed Einstein equations: scalar mode \label{pert_Einst}}

We are now in a position to specialize the
perturbed Einstein field equations~\eqref{einst_pert_gen}
to the case of scalar perturbations using the
linear combinations~\eqref{GandT_lincombo}:
\begin{subequations}\label{einst_pert2_lincombo}
\begin{xalignat}{2}
{\sf G}^{\Gamma}({}^{(1)}\!f) &=
3(1+w){}^{(1)}\!\Gamma; &\,
{\sf G}^{\Gamma}({}^{(2)}\!f) &=
3(1+w){}^{(2)}\!\Gamma - {\mathbb S}^{\Gamma},\label{eingamma}  \\
{\sf G}^{\pi}({}^{(1)}\!f)&=0; &\,
{\sf G}^{\pi}({}^{(2)}\!f) &=
-{\mathbb S}^{\pi},\label{einpi}  \\
{\sf G}^{q}({}^{(1)}\!f) &=
3(1+w){\cal H}{}^{(1)}\!V;  &\,
{\sf G}^{q}({}^{(2)}\!f)&=
3(1+w){\cal H}{}^{(2)}\!V - {\mathbb S}^{q}, \label{epl.4}\\
{\sf G}^{\rho}({}^{(1)}\!f)&= 3(1+w){}^{(1)}\!\bdelta ;  &\,
{\sf G}^{\rho}({}^{(2)}\!f)&=
3(1+w){}^{(2)}\!\bdelta - {\mathbb S}^{\rho}.\label{epl.3}
\end{xalignat}
where the complete source terms have the following form:
\begin{equation}  \label{einst_source_lincombo}
{\mathbb S} = {\mathbb G}({}^{(1)}\!f)-3(1+w){\mathbb T}({}^{(1)}\!f,{}^{(1)}\!M),
\end{equation}
\end{subequations}
for the superscripts ${\Gamma}, {\pi},q,\rho$.
In these equations the leading order terms ${\sf G}({}^{(1)}\!f)$ and
${\sf G}({}^{(2)}\!f)$ are given by~\eqref{lead_Gab} and
the source terms ${\mathbb G}({}^{(1)}\!f)$ and ${\mathbb T}({}^{(1)}\!f,{}^{(1)}\!M)$
are given by~\eqref{source_Gab} and~\eqref{source_Tab} in appendix~\ref{source}.

In  section~\ref{gov.eq} we will specialize
equations~\eqref{einst_pert2_lincombo} to the Poisson gauge
$(B=0)$, and label the remaining variables with a subscript ${}_{\mathrm p}$,
the uniform curvature gauge $(\psi=0)$, and label the remaining
variables with a subscript ${}_{\mathrm c}$
and the total matter gauge $(V=0)$, and label the remaining
variables with a subscript ${}_{\mathrm v}$.
When specifying a gauge we will use the following shorthand
notation for the source terms, for example in the Poisson gauge:
\begin{equation}  \label{einst_source_shorthand}
{\mathbb G}({}^{(1)}\!f_\mathrm{p}) = {\mathbb G}_\mathrm{p},
\qquad {\mathbb T}({}^{(1)}\!f_\mathrm{p},{}^{(1)}\!M_\mathrm{p}) = {\mathbb T}_\mathrm{p},
\end{equation}
for each of the superscripts $\Gamma, \pi, q, \rho,$ and similarly in the
other gauges.
%
%

The role played by
each of the four equations in the set~\eqref{einst_pert2_lincombo}
depends on the choice of gauge, as can be seen by referring to
the expressions~\eqref{lead_Gab} for the leading order terms (in what follows
we identify the four equations in~\eqref{einst_pert2_lincombo} by using the symbol for the
leading order Einstein tensor terms):
\begin{itemize}
\item[i)] The ${\sf G}^{\Gamma}$ equation gives
\begin{itemize}
\item[a)]
a second order evolution equation for $\psi_{\mathrm p}$ (the Bardeen equation)
in the Poisson gauge,
\item[b)] a first order evolution equation for $\phi_{\mathrm c}$ in the uniform curvature gauge.
\end{itemize}
\item[ii)] The ${\sf G}^{\pi}$ equation gives
\begin{itemize}
\item[a)] a constraint equation for $\phi_{\mathrm p}$ in the Poisson gauge,
\item[b)] a first order evolution equation for $B_{\mathrm c}$ in the uniform curvature gauge,
\item[c)] a first order evolution equation for $B_{\mathrm v}$ in the total matter gauge.
\end{itemize}
\item[iii)] The ${\sf G}^{q}$ equation gives
\begin{itemize}
\item[a)] a constraint equation for $V_{\mathrm p}$ in the Poisson gauge,
\item[b)] a constraint equation for $V_{\mathrm c}$ in the uniform curvature gauge,
\item[c)] a first order evolution equation for $\psi_{\mathrm v}$ in the total matter gauge.
\end{itemize}
\item[iv)] The ${\sf G}^{\rho}$ equation gives a constraint equation for ${\bdelta}$
in all three gauges.
\end{itemize}


Finally, before specializing the gauge, we present the general expression for
${}^{(r)}\!\bdelta,\,r=1,2$, valid in any temporal gauge,
that arises from the constraint referred to in iv) above, and
that will be useful later. Forming the (at first order gauge invariant) linear
combination~\eqref{epl.3}$-3$\eqref{epl.4} (\emph{i.e.}
the  ${\sf G}^{\rho} - 3{\sf G}^{q}$ equation)
and using the leading order term~\eqref{lG.4} we obtain
\begin{subequations} \label{delta_gen}
\begin{align}
{}^{(1)}\!\bdelta &= 3{\cal H}{}^{(1)}\!V +
\sfrac23(1+w)^{-1}{\cal H}^{-2}{\bf D}^2({}^{(1)}\!{\psi} - {\cal H}{}^{(1)}\!B),
\label{delta1_gen}\\
{}^{(2)}\!\bdelta &= 3{\cal H}{}^{(2)}\!V +
\sfrac13(1+w)^{-1}\left(2{\cal H}^{-2}{\bf D}^2({}^{(2)}\!{\psi} - {\cal H}{}^{(2)}\!B) +
{\mathbb S}^{\rho} - 3{\mathbb S}^q\right), \label{delta2_gen}
\end{align}
where
\begin{equation}
{\mathbb S}^{\rho} = {\mathbb G}^{\rho} - 3(1+w){\mathbb T}^{\rho}, \qquad
{\mathbb S}^q = {\mathbb G}^q - 3(1+w){\mathbb T}^q,  \label{delta2_source}
\end{equation}
\end{subequations}
using the notation in~\eqref{einst_source_lincombo}.

\section{The perturbed conservation equations \label{pert_cons}}

When using the perturbed Einstein equations,
the metric perturbations are determined by the evolution equations, while
the matter perturbations  are determined by the  constraint equations.
As an alternative approach one can use the perturbed conservation
equations to determine the evolution of the density and
velocity perturbations and use two of the perturbed Einstein
equations acting as constraints to determine the metric perturbations.

In order to determine the perturbed conservation equations
we associate an energy term $E(\epsilon)$ and a scalar
momentum term $M(\epsilon)={\cal S}^iE_i(\epsilon)$ with
the divergence $\bna\!_b(\epsilon)T^{b}\!_{a}(\epsilon)$ of the
stress-energy tensor $T^a\!_b(\epsilon)$ , where
\begin{subequations} \label{E_expansion}
\begin{align}
E(\epsilon) &= H^{-1}(\rho(\epsilon) + p(\epsilon))^{-1}
u^a(\epsilon)\bna\!_b(\epsilon)T^{b}\!_{a}(\epsilon), \\
E_i(\epsilon) &= (\rho(\epsilon) + p(\epsilon))^{-1}
\bna\!_b(\epsilon)T^{b}\!_{i}(\epsilon).
\end{align}
\end{subequations}
The Taylor expansion for $T^{b}\!_{a}(\epsilon)$ leads to
a Taylor series expansion for $E(\epsilon)$ :
\be E(\epsilon)= {}^{(0)}\!E +\epsilon\, {}^{(1)}\!E +\sfrac12 \epsilon^2\, {}^{(2)}\!E+...., \end{equation}
and similarly for $M(\epsilon)$. The coefficients in this expansion have a structure
analogous to the perturbations of the Einstein tensor and stress-energy tensor
in~\eqref{pert_Gab} and~\eqref{pert_Tab}:
\begin{subequations} \label{pert.cons}
\begin{alignat}{2}
{}^{(1)}\!E & = {\sf E}({}^{(1)}\!F), &\qquad
{}^{(1)}\!M &= {\sf M}({}^{(1)}\!F), \label{cons1} \\
{}^{(2)}\!E &= {\sf E}({}^{(2)}\!F) + {\mathbb E}({}^{(1)}\!F), &\qquad
{}^{(2)}\!M &= {\sf M}({}^{(2)}\!F) + {\mathbb M}({}^{(1)}\!F),  \label{cons2}
\end{alignat}
where
\begin{equation}
{F} = (\phi, {\cal H}{B},\psi, {\bdelta}, \Gamma, {\cal H}V).
\end{equation}
\end{subequations}
The first and second order perturbations
have a common \emph{leading order term} of the form ${\sf E}(F)$
or ${\sf M}(F)$ , where $F={}^{(1)}\!F$ or
$F={}^{(2)}\!F$  while ${}^{(2)}\!E$ and ${}^{(2)}\!M$
also have a \emph{source term} ${\mathbb E}({}^{(1)}\!F)$
and ${\mathbb M}({}^{(1)}\!F)$ which depends quadratically
on ${}^{(1)}\!F $.

Performing the perturbation expansion in~\eqref{E_expansion} to first order
gives the following expressions for the leading order terms:
\begin{subequations}\label{lead_divT}
\begin{align}
\mathsf{E}({F}) &= \partial_N(\bdelta - 3\psi) +
{\cal H}^{-2}{\bf D}^2({\cal H}{V} - {\cal H}{B}) + 3\Gamma , \label{E_lead_divT} \\
\mathsf{M}({F}) &= (\partial_N + 1 + q)({\cal H}V) +
\phi + c_s^2(\bdelta - 3{\cal H}{V}) +\Gamma,\label{M_lead_divT}
\end{align}
\end{subequations}
where we have used~\eqref{scale.H}, which introduces $q$ into the equation
and suggests that we use $1+q$ instead of $\sfrac32(1+w)$.
A similar but more lengthy calculation to second order
leads to the expressions for the quadratic source
terms ${\mathbb E}({}^{(1)}\!F)$ and ${\mathbb M}({}^{(1)}\!F)$
that are given by equations~\eqref{source_divT} in appendix~\ref{source}.

Referring to equation~\eqref{pert.cons}
the perturbed conservation equations at first and second order
are given by ${}^{(i)}\!E=0,\,i=1,2$, (conservation of energy) and
${}^{(i)}\!M=0,\,i=1,2$ (conservation of momentum).
From equations~\eqref{cons1} and~\eqref{cons2} we obtain
\begin{subequations} \label{pert.cons.final}
\begin{alignat}{2}
{\sf E}({}^{(1)}\!F)&=0, &\qquad
 {\sf M}({}^{(1)}\!F)&=0, \label{cons1.final} \\
 {\sf E}({}^{(2)}\!F) + {\mathbb E}({}^{(1)}\!F)&=0, &\qquad
{\sf M}({}^{(2)}\!F) + {\mathbb M}({}^{(1)}\!F)&=0,  \label{cons2.final}
\end{alignat}
\end{subequations}
where the leading order terms are given by~\eqref{lead_divT}
and the source terms are given by~\eqref{source_divT}.

The perturbed energy conservation equation at second order has been given
by Malik and Wands (2004)~\cite{malwan04} in the long wavelength limit
(see equation (5.33) where they use ${}^{(2)}\!\rho$ and ${}^{(2)}\!p$
rather than ${}^{(2)}\!{\bdelta}$  and ${}^{(2)}\!\Gamma$ as perturbation
variables).
We are aware of two general formulations of the perturbed conserved equations
to second order, namely, Hwang and Noh (2007)~\cite{hwanoh07a} and
Nakamura (2009)~\cite{nak09a}. For purposes of comparison we refer to their
equations when specialized to the case of a perfect fluid and scalar perturbations:
in~\cite{hwanoh07a} see equation (100) with (95) for conservation of energy,
and (101) for conservation of momentum, and in~\cite{nak09a} see
equations (4.8)-(4.10) for conservation of energy and (4.14), (4.18) and (4.19)
for conservation of momentum. In contrast to our approach these authors
use the unscaled density and pressure perturbations as matter variables
and do not introduce the non-adiabatic pressure perturbation $\Gamma$,
which means that an immediate comparison cannot be made. As regards gauge
choice, Hwang and Noh give their equations for an arbitrary choice of temporal
gauge, while Nakamura effectively uses the Poisson gauge.

\section{Ready-to-use systems of governing equations~\label{gov.eq}}

In this section, by specializing the perturbed
Einstein equations~\eqref{einst_pert2_lincombo} and
conservation equations~\eqref{lead_divT}
and~\eqref{pert.cons.final} to various gauges
we derive the ready-to-use systems of governing
equations described in the introduction.

\subsection{The Poisson gauge\label{poisson}}

The Poisson gauge is defined by the condition $B=0$. The scalar metric and
matter perturbations are denoted by ${\phi}_{\mathrm p}$, $\psi_{\mathrm p}$,
with $B_{\mathrm p}=0$, and ${V}_{\mathrm p}$, $\bdelta_{\mathrm p}$ with the
subscript ${}_\mathrm{p}$ indicating the Poisson gauge while a superscript
indicates the order of the perturbation, \emph{e.g.}
${}^{(r)}\!\psi_{\mathrm p},\, r=1,2$ (since ${}^{(1)}\!\Gamma$ and
${}^{(2)}\!\Gamma$ are gauge invariants, they will not have a subscript
in any gauge).

We insert $B=0$ into the leading order terms~\eqref{lead_Gab},
and label the remaining variables with a subscript
${}_{\mathrm p}$. These leading order terms (first and second order),
when inserted into equations~\eqref{einst_pert2_lincombo}, give
the perturbed Einstein equations in the Poisson gauge. It is convenient,
however, to obtain ${\bdelta}_{\mathrm p}$ directly by choosing the
Poisson gauge in equation~\eqref{delta_gen}. In addition, in order
to obtain the Bardeen equation~\eqref{p_gov2.1} below in a direct way
we form the linear combination~\eqref{eingamma} +
$2({\cal L}_1 + \sfrac13{\cal H}^{-2}{\bf D}^2)$\eqref{einpi}
of the perturbed Einstein equations and use the following relation  for
the leading order Einstein terms:
\begin{equation}
{\mathsf G}^{\Gamma}_{\mathrm p} + 2\left({\cal L}_1 +
\sfrac13{\cal H}^{-2}{\bf D}^2\right){\mathsf G}^{\pi}_{\mathrm p} =
2({\cal L}_\mathrm{B} - {\cal C}^2{\cal H}^{-2}{\bf D}^2)\psi_{\mathrm p},
\end{equation}
where we have used the linear combination
2\eqref{lG.1} + $2{\cal L}_1$\eqref{lG.2}.

\subsubsection{The Bardeen equation for $\psi_{\mathrm p}$}

At first order the above procedure leads to the following system:
\begin{subequations} \label{pert_gov1_p_simple}
\begin{align}
\left({\cal L}_B - c_s^2{\cal H}^{-2}{\bf D}^2 \right)
{}^{(1)}\!\psi_\mathrm{p} &= \sfrac32(1+w){}^{(1)}\!\Gamma ,\label{P1}\\
{}^{(1)}\!\phi_{\mathrm p} &={}^{(1)}\!\psi_{\mathrm p},\label{P4}  \\
{\cal H}{}^{(1)}\!{V}_{\mathrm p} &= -\sfrac23(1+w)^{-1}(\partial_N {}^{(1)}\!\psi_{\mathrm p} +
{}^{(1)}\!\phi_{\mathrm p}),\label{P2} \\
{}^{(1)}\!{\bdelta}_{\mathrm p} &= 3{\cal H}{}^{(1)}\!{V}_{\mathrm p} +
\sfrac23(1+w)^{-1}{\cal H}^{-2}{\bf D}^2\,{}^{(1)}\!\psi _{\mathrm p},\label{P3}
\end{align}
\end{subequations}
where ${\cal L}_B$ is given
by~\eqref{LB_expand}, although the product
form~\eqref{LB} of the operator is useful when solving the equation.
Observe that ${}^{(1)}\!\psi_\mathrm{p}$ is the primary dynamical variable
and is determined by the Bardeen equation~\eqref{P1}.

The second order perturbation equations have the following form:
\begin{subequations} \label{p_gov2}
\begin{align}
\left({\cal L}_B - c_s^2{\cal H}^{-2}{\bf D}^2 \right) {}^{(2)}\!\psi_\mathrm{p} &=
\sfrac32(1+w){}^{(2)}\!\Gamma -
\sfrac12 {\mathbb S}^{\Gamma}_\mathrm{p} -
({\cal L}_1 + \sfrac13{\cal H}^{-2}{\bf D}^2 ){\mathbb S}^{\pi}_{\mathrm p}, \label{p_gov2.1}  \\
{}^{(2)}\!\phi_{\mathrm p} &= {}^{(2)}\!\psi_{\mathrm p} +
{\mathbb S}^{\pi}_{\mathrm p}, \label{p_gov2.2} \\
{\cal H}{}^{(2)}\!{V}_{\mathrm p} &=
-\sfrac23(1+w)^{-1}\!\left(\partial_N {}^{(2)}\!\psi_{\mathrm p} +
{}^{(2)}\!\phi_{\mathrm p} - \sfrac12{\mathbb S}^q _{\mathrm p}\right),
\label{p_gov2.4}\\
{}^{(2)}\!{\bdelta}_{\mathrm p} &=  3 {\cal H}{}^{(2)}\!{V}_{\mathrm p} +
\sfrac23(1+w)^{-1}\!\left({\cal H}^{-2}{\bf D}^2{}^{(2)}\!{\psi}_{\mathrm p} +
\sfrac12{\mathbb S}^{\rho}_{\mathrm p} - \sfrac32{\mathbb S}^q_{\mathrm p}\right),\label{bdelta2_p}
\end{align}
\end{subequations}
where the source terms ${\mathbb S}_{\mathrm p}$, for the superscripts
${\Gamma}, {\pi},q,\rho$, are given by
\begin{equation}  \label{source_p}
{\mathbb S}_{\mathrm p}={\mathbb G}_{\mathrm p}-3(1+w){\mathbb T}_{\mathrm p},
\end{equation}
using the notation~\eqref{einst_source_lincombo} and~\eqref{einst_source_shorthand}.
To complete the specification of the equations we need to give the explicit form of
the source terms ${\mathbb G}_{\mathrm p}$ and
${\mathbb T}_{\mathrm p}$, which are obtained by specializing
equations~\eqref{source_Gab} and~\eqref{source_Tab}
in appendix~\ref{source} to the Poisson gauge ($B=0$) and inserting
the relations ${}^{(1)}\!\phi_{\mathrm p} = {}^{(1)}\!\psi_{\mathrm p}$
(equation~\eqref{P4}). Equations~\eqref{source_Gab} yield\footnote{These expressions
have been given by Uggla and Wainwright (2013)~\cite{uggwai13a}, see equation (35).
Here and elsewhere, in order to simplify the notation
 we omit the superscript ${}^{(1)}$ on the linear perturbations
in the source terms.}
\begin{subequations}\label{source_ein_p}
\begin{align}
{\mathbb G}^{\Gamma}_{\mathrm p} &= - 8{\cal L}_1(\psi_{\mathrm p}^2) +
\sfrac23\!\left(1 + 3c_s^2\right){\mathbb X}_{\mathrm p} -
\sfrac83{\cal H}^{-2}({\bf D}\psi_{\mathrm p})^2,\\
{\mathbb G}^{\pi}_{\mathrm p} &= 4\!\left(\psi_{\mathrm p}^2 -
{\mathbb D}_0(\psi_{\mathrm p})\right),  \\
{\mathbb G}^q_{\mathrm p} &= 4\left(2\psi_{\mathrm p}^2 -
{\cal S}^i\!\left[(\partial_N\psi_{\mathrm p}){\bf D}_i\psi_{\mathrm p}\right]\right),\\
{\mathbb G}^{\rho}_{\mathrm p} &= 24\psi_{\mathrm p}^2 - 2{\mathbb X}_{\mathrm p},
\end{align}
where the mode extraction operator ${\cal S}^i$ was given in~\eqref{modeextractop}
while the spatial differential operator ${\mathbb D}_0$ is defined
in equation~\eqref{QD2} in appendix~\ref{source}.
In addition,
\begin{equation} \label{cal_R}
{\mathbb X}_{\mathrm p} = - 3(\partial_N \psi_{\mathrm p})^2 +
5{\cal H}^{-2}({\bf D}\psi_{\mathrm p})^2 - 4{\cal H}^{-2}{\bf D}^2 \!\psi_{\mathrm p}^2.
\end{equation}
\end{subequations}
Equations~\eqref{source_Tab} yield:
\begin{subequations}  \label{source_Tab_p}
\begin{align}
{\mathbb T}^{\Gamma}_{\mathrm p} &= \sfrac23(1-3c_s^2)({\bf D} V_{\mathrm p})^2
- \sfrac13(\partial_N c_s^2)(\bdelta_{\mathrm p})^2 -
\sfrac23 {}^{(1)}{\bdelta_{\mathrm p}}\left(\partial_N - 3(1+c_s^2)\right){\Gamma}, \\
{\mathbb T}^{\pi}_{\mathrm p} &= 2{\mathbb D}_0({\cal H}V_{\mathrm p}),  \\
{\mathbb T}^q_{\mathrm p} &=
{\cal S}^i\left[2\left((1+c_s^2)\bdelta_{\mathrm p} -
\phi_{\mathrm p} + \Gamma\right){\bf D}_i({\cal H}V_{\mathrm p})\right],\\
{\mathbb T}^{\rho}_{\mathrm p} &= 2({\bf D} V_{\mathrm p})^2.
\end{align}
\end{subequations}
The perturbed Einstein
equations at second order in the Poisson gauge have been given in different
forms by various authors.\footnote{See, for example, Noh and Hwang (2004)~\cite{nohhwa04},
equation (303), and Nakamura (2007)~\cite{nak07}, equations (6.38), (6.41), (6.42) and (6.44).}

\subsubsection{Coupled evolution equations for $\psi_{\mathrm p}$ and
 $V_{\mathrm p}$}

An alternative approach to analyzing the dynamics in the Poisson
gauge is to use $\psi_{\mathrm p}$ and $V_{\mathrm p}$ as
primary dynamical variables, with the perturbed Einstein
equation ${\mathsf G}^q_{\mathrm p}$ as evolution equation for $\psi_{\mathrm p}$
and the perturbed conservation of momentum
equation ${\mathsf M}_{\mathrm p}$ as evolution equation for $V_{\mathrm p}$.

To obtain the first equation we use the perturbed Einstein equation~\eqref{p_gov2.4},
with ${}^{(2)}\!\phi_{\mathrm p}$
eliminated using~\eqref{p_gov2.2}.
To obtain the second equation we use the perturbed conservation of momentum
equation~\eqref{cons2.final} at second order in the Poisson gauge which
reads
\begin{equation}
(\partial_N + 1 + q)({\cal H}{}^{(2)}\!V_{\mathrm p}) +
{}^{(2)}\!\phi_{\mathrm p} + c_s^2({}^{(2)}\!\bdelta - 3{\cal H}{}^{(2)}\!V_{\mathrm p}) +
{}^{(2)}\!\Gamma +{\mathbb M}_{\mathrm p}=0,
\end{equation}
and use~\eqref{p_gov2.2} to eliminate ${}^{(2)}\!\phi_{\mathrm p}$ and~\eqref{bdelta2_p}
to eliminate ${}^{(2)}\!\bdelta_{\mathrm p}-3{\cal H}{}^{(2)}\!V_{\mathrm p}$.
The resulting equations are as follows:
\begin{subequations} \label{evol2_psi,V}
\begin{align}
(\partial_N +1){}^{(2)}\!\psi_{\mathrm p}+(1+q){\cal H}{}^{(2)}\!V_{\mathrm p}
-\sfrac12 {\mathbb S}^{q}_{\mathrm p}+ {\mathbb S}^{\pi}_{\mathrm p} &= 0,\\
(\partial_N +1 +q)({\cal H}{}^{(2)}\!V_{\mathrm p}) +
\left(1+(1+q)^{-1}c_s^2{\cal H}^{-2}{\bf D}^2 \right){}^{(2)}\!\psi_{\mathrm p}
+{}^{(2)}\!\Gamma+ {\mathbb S} &= 0,
\end{align}
where
\begin{equation}
{\mathbb S} ={\mathbb M_{\mathrm p}} + {\mathbb S^{\pi}_{\mathrm p}} +
\sfrac12(1+q)^{-1}c_s^2 ({\mathbb S^{\rho}_{\mathrm p}} -3{\mathbb S^{q}_{\mathrm p}}).
\end{equation}
\end{subequations}
The source terms with kernel $\mathbb S_{\mathrm p}$
are given by~\eqref{source_p} and  ${\mathbb M_{\mathrm p}}$ is given
by~\eqref{M_source}.
Equations~\eqref{evol2_psi,V} form a coupled system of evolution equations
for ${}^{(2)}\!\psi_{\mathrm p}$ and ${}^{(2)}\!V_{\mathrm p}$. The corresponding
system for ${}^{(1)}\!\psi_{\mathrm p}$ and ${}^{(1)}\!V_{\mathrm p}$ is
obtained by dropping the source terms and changing ${}^{(2)}$ to ${}^{(1)}$.

The system of equations~\eqref{evol2_psi,V} has the same dynamical content
as the second order Bardeen equation~\eqref{p_gov2.1} for ${}^{(2)}\!\psi_{\mathrm p}$.
One can derive the Bardeen equation from~\eqref{evol2_psi,V} by solving the first equation
algebraically for ${\cal H}{}^{(2)}\!V_{\mathrm p}$ and substituting it into the second
equation. The difference is that the source term obtained in this way has a different form
from the source term in~\eqref{p_gov2.1}.

\subsection{The uniform curvature gauge\label{ucg}}

The uniform curvature gauge is defined by the condition $\psi=0$. The scalar
metric perturbations are denoted by $\phi_{\mathrm c}$, ${B}_{\mathrm c}$,
with $\psi_{\mathrm c}=0$, and the matter variables by ${\bdelta}_{\mathrm c}$,
${V}_{\mathrm c}$, and $\Gamma$, with a superscript indicating the order of the
perturbation, \emph{e.g.} ${}^{(r)}\!\phi_{\mathrm c}$, $r=1,2$.
We insert $\psi=0$ into the leading order terms~\eqref{lead_Gab},
and label the remaining variables with a subscript
${}_{\mathrm c}$. These leading order terms (first and second order),
when inserted into equations~\eqref{einst_pert2_lincombo}, give
the perturbed Einstein equations in the uniform curvature gauge. It is convenient,
however, to obtain ${\bdelta}_{\mathrm c}$ directly by choosing the
uniform curvature gauge in equation~\eqref{delta_gen}.

\subsubsection{Coupled evolution equations for $\phi_{\mathrm c}$ and
 $B_{\mathrm c}$}

At first order the above procedure leads to the following system:
\begin{subequations}\label{ucg_gov1}
\begin{align}
(1+q)\partial_N((1+q)^{-1} {}^{(1)}\!{\phi_\mathrm{c}}) &= -c_s^2{\cal H}^{-2}{\bf D}^2({\cal H}{}^{(1)}\!{B}_{\mathrm c}) + (1+q){}^{(1)}\Gamma, \label{ucg_gov1.1} \\
\partial_N(a^2\,{}^{(1)}\!B_{\mathrm c} )&= -a^2 {\cal H}^ {-1}{}^{(1)}\!{\phi_\mathrm{c}},  \label{ucg_gov1.2}  \\
{\cal H}{}^{(1)}\!{V}_{\mathrm c} &=  -(1+q)^{-1} {}^{(1)}\!{\phi_\mathrm{c}},\label{ucg_gov1.4}\\
{}^{(1)}\!{\bdelta}_{\mathrm c} &= 3 {\cal H}{}^{(1)}\!{V}_{\mathrm c} -
(1+q)^{-1}{\cal H}^{-2}{\bf D}^2 ({\cal H}{}^{(1)}\!{B}_{\mathrm c}),  \label{ucg_gov1.3}
\end{align}
\end{subequations}
where we have used the expression ${\cal L}_1f = (1+q)\partial_N\left((1+q)^{-1} f\right)$,
given in~\eqref{L1}, which introduces $q$ into the equations. We have chosen not
to replace $1+q$ by $\sfrac32(1+w)$. At second order we obtain:
\begin{subequations} \label{ucg_gov2}
\begin{align}
\begin{split}
(1+q)\partial_N((1+q)^{-1} {}^{(2)}\!{\phi_\mathrm{c}}) &=
- c_s^2{\cal H}^{-2}{\bf D}^2( {\cal H}{}^{(2)}\!{B}_{\mathrm c}) +
(1+q){}^{(2)}\Gamma \\
& \qquad \qquad \qquad \qquad - \sfrac12{\mathbb S}^{\Gamma}_{\mathrm c} -
\sfrac13{\cal H}^{-2}{\bf D}^2{\mathbb S}^{\pi}_{\mathrm c},
\end{split}
\label{ucg_gov2.1}  \\
\partial_N(a^2\,{}^{(2)}\!B_{\mathrm c} ) &=
- a^2 {\cal H}^ {-1}\left({}^{(2)}\!{\phi_\mathrm{c}}
- {\mathbb S}^{\pi}_{\mathrm c}\right),  \label{ucg_gov2.2} \\
{\cal H}{}^{(2)}\!{V}_{\mathrm c} &=  - (1+q)^{-1}({}^{(2)}\!{\phi_\mathrm{c}} -
\sfrac12 {\mathbb S}^q_{\mathrm c}),  \label{ucg_gov2.4}\\
{}^{(2)}\!{\bdelta}_{\mathrm c} &=  3{\cal H}{}^{(2)}\!{V}_{\mathrm c} +
(1+q)^{-1}\!\left(-{\cal H}^{-2}{\bf D}^2({\cal H}{}^{(2)}\!{B}_{\mathrm c}) +
\sfrac12 ({\mathbb S}^{\rho}_{\mathrm c} -3{\mathbb S}^q_{\mathrm c})\right).  \label{bdelta2_uc}
\end{align}
\end{subequations}
The source terms with kernel $\mathbb S_{\mathrm c}$
are given by
\begin{equation}  \label{source_c}
{\mathbb S}_{\mathrm c}={\mathbb G}_{\mathrm c}-3(1+w){\mathbb T}_{\mathrm c},
\end{equation}
using the notation~\eqref{einst_source_lincombo} and~\eqref{einst_source_shorthand}.
The source terms for the Einstein tensor, with kernel ${\mathbb G}_{\mathrm c}$, are given by
equations~\eqref{source_Gab} in appendix~\ref{source} with
$\psi_{\mathrm c}=0$:\footnote{The source terms for
the perturbed Einstein tensor in the uniform
curvature gauge have been given by Uggla and Wainwright (2013)~\cite{uggwai13a}
(see equations (96)).}
\begin{subequations}  \label{G_source_c}
\begin{align}
\begin{split}
{\mathbb G}^{\Gamma}_{\mathrm c} &=
-2{\cal L}_1(4\phi_{\mathrm c}^2  -({\bf D}{B}_{\mathrm c})^2 )
- \sfrac43 {\cal H}^{-2}\!\left[(\partial_N\phi_{\mathrm c} -
4\phi_{\mathrm c}){\bf D}^2({\cal H}B_{\mathrm c})
+ ({\bf D}\phi_{\mathrm c})^2\right]  \\
& \quad -\sfrac13(2+3c_s^2){\mathbb W}_{\mathrm c}
- \sfrac23(1 + 3c_s^2){\cal H}^{-2}{\bf D}^2{\mathbb D}_2({B}_{\mathrm c}) ,\end{split}  \label{red_source.1}\\
{\mathbb G}^{\pi}_{\mathrm c} &= 2{\mathbb D}_0( \phi_{\mathrm c}) +
2{\cal S}^{ij}[(2({\bf D}_{ i}\phi_{\mathrm c}){\bf D}_{j}({\cal H}{B}_{\mathrm c}) +
\left(\partial_N \phi_{\mathrm c} \right){\bf D}_{ij}({\cal H}{B}_{\mathrm c})] +
{\mathbb D}_2({B}_{\mathrm c}), \\
{\mathbb G}^q_{\mathrm c} &=   2(4\phi_{\mathrm c}^2 - ({\bf D}{B}_{\mathrm c})^2)^2 -
2{\cal H}^{-2}{\cal S}^i [({\bf D}_j\phi_{\mathrm c})\,({\bf D}^j\!_i -
\sfrac23\delta^j\!_i{\bf D}^2)({\cal H}{B}_{\mathrm c})],\\
{\mathbb G}^{\rho}_{\mathrm c} &= 6(4\phi_{\mathrm c}^2  - ({\bf D}{B}_{\mathrm c})^2)^2 +
{\mathbb W}_{\mathrm c} + 2{\cal H}^{-2}{\bf D}^2{\mathbb D}_2({B}_{\mathrm c}),
\end{align}
where
\begin{equation}
{\mathbb W}_{\mathrm c} =
{\cal H}^{-2}\left[8\phi_{\mathrm c}{\bf D}^2({\cal H}{B}_{\mathrm c}) +
4({\bf D}^k\phi_{\mathrm c}){\bf D}_k({\cal H}{B}_{\mathrm c})\right].
\end{equation}
\end{subequations}
The spatial differential operators ${\mathbb D}_0$
and ${\mathbb D}_2$ are defined in equations~\eqref{QD2} and~\eqref{QD3} in appendix~\ref{source}.

The source terms for the stress-energy tensor, with kernel ${\mathbb T}$, are given by
equations~\eqref{source_Tab} in appendix~\ref{source},
specialized to the uniform curvature gauge:
\begin{subequations}  \label{T_source_c}
\begin{align}
{\mathbb T}^{\rho}_{\mathrm c} &= \gamma^{ij}({\mathbb V}_{2,{\mathrm c}})_{ij},  \\
{\mathbb T}^{\Gamma}_{\mathrm c} &= \sfrac13(1-3c_s^2)\gamma^{ij}({\mathbb V}_{2,{\mathrm c}})_{ij}
- \sfrac13(\partial_N c_s^2)\bdelta_{\mathrm c}^2 -
\sfrac23 {\bdelta}_{\mathrm c}\left(\partial_N - 3(1+c_s^2) \right)\!{\Gamma}, \label{T_gam} \\
{\mathbb T}^q_{\mathrm c} &= 2{\cal S}^i\left[\left((1+c_s^2)\bdelta_{\mathrm c} - \phi_{\mathrm c} +
\Gamma\right){\bf D}_i({\cal H}{V_{\mathrm c}})\right], \\
{\mathbb T}^{\pi}_{\mathrm c} &= {\cal H}^2{\cal S}^{ij}({\mathbb V}_{2,{\mathrm c}})_{ij},
\end{align}
where
\begin{equation} \label{Vij_c}
({\mathbb V}_{2,{\mathrm c}})_{ij} =2{\cal H}^{-2}
({\bf D}_i{\cal H}{V_{\mathrm c}}){\bf D}_j\!\left({\cal H}{V_{\mathrm c}} -
{\cal H}{B_{\mathrm c}}\right).
\end{equation}
\end{subequations}
To the best of our knowledge the system of equations~\eqref{ucg_gov2}
and the associated source terms
are new. We comment  on the utility of  these equations in the discussion
in section~\ref{discussion}.

\subsection{The total matter gauge  \label{tmg}}

The total matter gauge is defined by the condition $V=0$. There are thus three
metric perturbations variables, $\phi_{\mathrm v}$, $\psi_{\mathrm v}$ and $B_{\mathrm v}$,
but only two matter perturbation variables ${\bdelta}_{\mathrm v}$ and $\Gamma$.
The perturbations of the stress-energy tensor thereby simplify. It follows from~\eqref{GandT_lincombo}
and~\eqref{source_Tab} that the leading order terms and the source terms, respectively, satisfy
\begin{equation}\label{Tab_comov}
{\sf T}^q_{\mathrm v} = 0, \qquad {\sf T}^{\pi}_{\mathrm v} = 0; \qquad
{\mathbb T}^{\rho}_{\mathrm v} = 0, \qquad {\mathbb T}^q_{\mathrm v} = 0, \qquad
{\mathbb T}^{\pi}_{\mathrm v} = 0.
\end{equation}
%

\subsubsection{Coupled evolution equations for $\psi_{\mathrm v}$ and
 $B_{\mathrm v}$}

When working in the total matter gauge it is convenient to replace the perturbed Einstein
equation associated with ${\sf G}^{\Gamma}$ by the perturbed conservation of momentum
equation, since this equation determines $\phi_{\mathrm v}$ algebraically. Specifically,
in the total matter gauge the perturbed conservation of
momentum equations~\eqref{M_lead_divT} and~\eqref{M_source} lead to
\begin{subequations} \label{phi_v}
\begin{align}
{}^{(1)}\!\phi_{\mathrm v} &= -c_s^2 {}^{(1)}\!\bdelta_{\mathrm v}  - {}^{(1)}\!\Gamma, \\
{}^{(2)}\!\phi_{\mathrm v} &= -c_s^2 {}^{(2)}\!\bdelta_{\mathrm v}  -
{}^{(2)}\!\Gamma - {\mathbb M}_{\mathrm v}, \label{phi_v2}
\end{align}
where
\begin{equation}
{\mathbb M}_{\mathrm v} = - 2\phi_{\mathrm v}^2 + ({\bf D}{B}_{\mathrm v})^2 - [c_s^2(1+c_s^2)
+ \sfrac13\partial_N c_s^2]\bdelta_{\mathrm v}^2 -\Gamma^2 -
\sfrac23{\bdelta}_{\mathrm v} \partial_{N}\Gamma +
2{\cal S}^i[\Gamma {\bf D}_i{\bdelta}_{\mathrm v}]. \label{source_M}
\end{equation}
\end{subequations}
On substituting~\eqref{Tab_comov} into the perturbed Einstein equations~\eqref{einst_pert2_lincombo} the first and second
order equations, excluding the ${\sf G}^{\Gamma}$ equation, become
\begin{subequations}\label{einst_pert1_lincombo_comov}
\begin{xalignat}{2}
{\sf G}^{\pi}({}^{(1)}\!f_{\mathrm v}) &= 0, &\quad {\sf G}^{\pi}({}^{(2)}\!f_{\mathrm v}) + {\mathbb G}^{\pi}_{\mathrm v} &= 0, \\
{\sf G}^{q}({}^{(1)}\!f_{\mathrm v}) &= 0, &\quad
{\sf G}^{q}({}^{(2)}\!f_{\mathrm v}) + {\mathbb G}^{q}_{\mathrm v} &= 0, \\
{\sf G}^{\rho}({}^{(1)}\!f_{\mathrm v})&= 3(1+w){}^{(1)}\!\bdelta_{\mathrm v}, &\quad
{\sf G}^{\rho}({}^{(2)}\!f_{\mathrm v}) + {\mathbb G}^{\rho}_{\mathrm v} &=
3(1+w){}^{(2)}\!\bdelta_{\mathrm v}.
\end{xalignat}
\end{subequations}

As governing equations we use equations~\eqref{phi_v} in conjunction with
the perturbed Einstein equations~\eqref{einst_pert1_lincombo_comov}.
To obtain the detailed form of the equations we substitute the expressions for the leading
terms of the Einstein tensor from equations~\eqref{lead_Gab}.
At first order we obtain
\begin{subequations}  \label{totmat_1}
\begin{align}
\partial_N {}^{(1)}\!{\psi}_{\mathrm v}
 &= c_s^2 {}^{(1)}\!\bdelta_{\mathrm v}
 + {}^{(1)}\!\Gamma, \label{totmat_1.2}   \\
{\cal H}a^{-2}\partial_N(a^2\,{}^{(1)}\!B_{\mathrm v} ) &=
{}^{(1)}\!{\psi}_{\mathrm v} + c_s^2 {}^{(1)}\!\bdelta_{\mathrm v}
 + {}^{(1)}\!\Gamma ,  \label{totmat_1.3}    \\
{}^{(1)}\!{\bdelta}_{\mathrm v} &=
\sfrac23(1+w)^{-1}{\cal H}^{-2}{\bf D}^2 ({}^{(1)}\!{\psi}_{\mathrm v} -
{\cal H}{}^{(1)}\!{B}_{\mathrm v}), \label{totmat_1.4}
\end{align}
\end{subequations}
while the second order equations can be written as
\begin{subequations}\label{totmat_2}
\begin{align}
\partial_N {}^{(2)}\!{\psi}_{\mathrm v}  &=
c_s^2 {}^{(2)}\!\bdelta_{\mathrm v}+
{}^{(2)}\!\Gamma  + {\mathbb M}_{\mathrm v} +
\sfrac12{\mathbb G}^q_{\mathrm v},  \label{totmat_2.2}  \\
{\cal H}a^{-2}\partial_N(a^2\,{}^{(2)}\!B_{\mathrm v} ) &= {}^{(2)}\!{\psi}_{\mathrm v}+
c_s^2 {}^{(2)}\!\bdelta_{\mathrm v}+
{}^{(2)}\!\Gamma  + {\mathbb M}_{\mathrm v}
+ {\mathbb G}^{\pi}_{\mathrm v},   \label{totmat_2.3}   \\
{}^{(2)}\!{\bdelta}_{\mathrm v} &=
\sfrac23(1+w)^{-1}\left({\cal H}^{-2}{\bf D}^2 ({}^{(2)}\!{\psi}_{\mathrm v} -
{\cal H}{}^{(2)}\!{B}_{\mathrm v}) +
\sfrac12({\mathbb G}_{\mathrm v}^{\rho} -3{\mathbb G}_{\mathrm v}^q)\right), \label{totmat_2.4}
\end{align}
\end{subequations}
where ${\mathbb M}_{\mathrm v}$ is given by~\eqref{source_M}.
The Einstein source terms, labelled ${\mathbb G}_{\mathrm v}$, are
obtained by evaluating equation~\eqref{source_Gab} in appendix~\ref{source}
in the total matter gauge. Since the
three metric perturbations are in general non-zero in this gauge
the Einstein source terms do not simplify in general.
However, as we will explain in the Discussion,
in the two benchmark problems mentioned in
the Introduction, additional restrictions  arise which simplify the leading
order Einstein terms and the
Einstein source terms significantly, making the total matter gauge an ideal choice
for these problems.

\subsection{The perturbed conservation equations}
In this subsection we use the matter variables $\bdelta$ and
$V$ as the primary dynamical variables and use both
perturbed conservation equations (energy and momentum)
to obtain the evolution equations.

The perturbed conservation equations at second order are given
by equations~\eqref{lead_divT} and~\eqref{cons2.final}, which we repeat here:
\begin{subequations}  \label{cons2.general}
\begin{align}
{}^{(2)}\!E &= \partial_N({}^{(2)}\!\bdelta - 3{}^{(2)}\!\psi) +
{\cal H}^{-2}{\bf D}^2({\cal H}{}^{(2)}\!{V} - {\cal H}{}^{(2)}\!{B}) + 3{}^{(2)}\!\Gamma + {\mathbb E} =0,\\
{}^{(2)}\!M &= (\partial_N + 1 + q)({\cal H}{}^{(2)}\!V) +
{}^{(2)}\!\phi + c_s^2({}^{(2)}\!\bdelta - 3{\cal H}{}^{(2)}\!{V}) + {}^{(2)}\!\Gamma +{\mathbb M}=0.
\end{align}
\end{subequations}
These equations provide evolution equations for $\bdelta$ and
$V$, but they do not form a closed evolution system since they are coupled
to the metric perturbations. However we can circumvent this difficulty by
an appropriate use of two gauges: in the following section
we consider $\bdelta_{\mathrm v}$
(total matter gauge) and $V_{\mathrm p}$ (Poisson gauge).

\subsubsection{Coupled evolution equations for $\bdelta_{\mathrm v}$ and
 ${\bf D}^2 V_{\mathrm p}$ \label{b_v.V_p} }

To obtain the first equation we calculate
${}^{(2)}\!E_{\mathrm v} - 3{}^{(2)}\!M_{\mathrm v}$
starting with~\eqref{cons2.general}. We eliminate $\partial_N {}^{(2)}\!\psi_{\mathrm v}$
using
\begin{equation}
\partial_N {}^{(2)}\!\psi_{\mathrm v} +{}^{(2)}\!\phi_{\mathrm v}=
\sfrac12\mathbb{G}_{\mathrm v}^{q},
\end{equation}
which follows from equations~\eqref{phi_v2} and~\eqref{totmat_2.2}, and we replace
${}^{(2)}\!B_{\mathrm v}$ by $-{}^{(2)}\!V_{\mathrm p}$ plus source terms using
the change of gauge formula
\begin{equation}
{\cal H}{}^{(2)}\!B_{\mathrm v} +{\cal H}{}^{(2)}\!V_{\mathrm p}=
{\mathbb S}[B_{\mathrm v} + V_{\mathrm p}],
\end{equation}
where the source term is given by~\eqref{B+V.source}. Here we have introduced
the notation ${\mathbb S}[\dots]$ for the source terms associated with
a change of gauge formula at second order.
To obtain the second equation we evaluate ${\bf D}^2 {}^{(1)}\!M_{\mathrm p}$
starting with~\eqref{cons2.general}.
We replace ${}^{(2)}\!\phi_{\mathrm p}$ by ${}^{(2)}\!\psi_{\mathrm p}$
plus source terms using~\eqref{p_gov2.2}. We next use the GR version of
Poisson's equation at second order,
\begin{equation} \label{M.2}
(1+q){}^{(2)}\!\bdelta_{\mathrm v} - {\cal H}^{-2}{\bf D}^2{}^{(2)}\!\psi_{\mathrm p} =
{\mathbb S}_{Poisson},
\end{equation}
where the source term is given by~\eqref{S_Poisson},
to express ${\bf D}^2{}^{(2)}\!\psi_{\mathrm p}$ in terms of
${}^{(2)}\!\bdelta_{\mathrm v}$ plus source terms. We finally express
${}^{(2)}\!{\bdelta}_{\mathrm p} - 3{}^{(2)}\!{\cal H}V_{\mathrm p}$ in terms
of ${}^{(2)}\!{\bdelta}_{\mathrm v}$ and source terms using the change of
gauge formula
\begin{equation}\label{M.3}
{}^{(2)}\!{\bdelta}_{\mathrm v} -  {}^{(2)}\!{\bdelta}_{\mathrm p} +
3{}^{(2)}\!{\cal H}V_{\mathrm p}=
{\mathbb S}[{\bdelta}_{\mathrm v} - {\bdelta}_{\mathrm p} + {\cal H}V_{\mathrm p}],\end{equation}
where the source term is given by~\eqref{delta_p,v_source}.
This procedure leads to the following system of evolution equations:
\begin{subequations} \label{delta_v,V_p_evol2}
\begin{equation}
(\partial_N - 3c_s^2){}^{(2)}\!\bdelta_{\mathrm v} +
{\cal H}^{-2}{\bf D}^2({\cal H}{}^{(2)}\!{V}_{\mathrm p}) + {\mathbb S}_{\rho}=0,
\end{equation}
\begin{equation}
(\partial_N + 1 - q)[{\cal H}^{-2}{\bf D}^2({\cal H}{}^{(2)}\!V_{\mathrm p})] +
(1 + q + c_s^2{\cal H}^{-2}{\bf D}^2){}^{(2)}\!\bdelta_{\mathrm v}
+ {\cal H}^{-2}{\bf D}^2\,{}^{(2)}\!\Gamma + {\mathbb S}_{V} = 0,
\end{equation}
\end{subequations}
where the source terms are
\begin{subequations} \label{source_general}
\begin{align}
{\mathbb S}_{\rho}&= \mathbb{E}_{\mathrm v} -
3\mathbb{M}_{\mathrm v}
 -\sfrac32 \mathbb{G}_{\mathrm v}^q -
 {\cal H}^{-2}{\bf D}^2({\mathbb S}[B_{\mathrm v} + V_{\mathrm p}]) ,
 \label{source_rho} \\
{\mathbb S}_{V}&= {\cal H}^{-2}{\bf D}^2\left(\mathbb{M}_{\mathrm p} +
\mathbb{S}_{\mathrm p}^{\pi}
+ c_s^2\,{\mathbb S}[\bdelta_{\mathrm p} -
3{\cal H}V_{\mathrm p} - {\bdelta}_{\mathrm v}] \right)
- {\mathbb S}_{Poisson}. \label{source_V}
\end{align}
\end{subequations}
The source terms $\mathbb{E}_{\mathrm v},\mathbb{M}_{\mathrm v}$
and $\mathbb{M}_{\mathrm p}$ are obtained by choosing the
total matter gauge and the Poisson gauge in the general
formulas~\eqref{source_divT} in appendix~\ref{source}, and the source terms associated with
a change of gauge formulas are given by~\eqref{gauge.change.source}.
The source term ${\mathbb S}_{Poisson}$, which is
the source term in the general relativity (GR) version of Poisson's equation~\eqref{M.2}
at second order, requires explanation. This equation is derived by
choosing the Poisson gauge in equation~\eqref{delta2_gen}
and then using~\eqref{M.3} to express $\bdelta_{\mathrm p}$
in terms of $\bdelta_{\mathrm v}$. Collecting all the source terms
gives the expression for ${\mathbb S}_{Poisson}$ in equation~\eqref{S_Poisson}.

\subsubsection{Second order evolution equation for $\bdelta_{\mathrm v}$
\label{evol.delta_v} }

In this section the first order evolution equations for
$\bdelta_{\mathrm v}$ and ${V}_{\mathrm p}$ are combined to give a second order
evolution equation for $\bdelta_{\mathrm v}$.
This evolution equation for the density perturbation  is
obtained by eliminating ${\bf D}^2({}^{(2)}\!{V}_{\mathrm p})$ in
equations~\eqref{delta_v,V_p_evol2}. After expanding the derivatives
we obtain
\begin{subequations}  \label{delta_v,evol,general}
\begin{equation}
({\cal L}_D  - c_s^2 {\cal H}^{-2}{\bf D}^2){}^{(2)}\!\bdelta_{\mathrm v} + {\mathbb S}_{{\cal L}_D} =
{\cal H}^{-2}{\bf D}^2\,{}^{(2)}\!\Gamma,
\end{equation}
where the differential operator ${\cal L}_D$ is given by
\begin{equation}
{\cal L}_D = \partial _N^2+
(1-q-3c_s^2)\partial _N - (1-3c_s^2)(1+q)-6c_s^2-3\partial_N c_s^2,
\end{equation}
and the source term ${\mathbb S}_{{\cal L}_D}$ is given by
\begin{equation} \label{delta_v,source1}
{\mathbb S}_{{\cal L}_D}= (\partial_N+1+q){\mathbb S}_{\rho} -
{\mathbb S}_{V}.
\end{equation}
\end{subequations}
We note in passing that the source term in the
evolution equation~\eqref{delta_v,evol,general}
simplifies significantly in the case of a perturbed $\Lambda CDM$
universe, which permits one to quickly find the second order density perturbation
${}^{(2)}\!\bdelta_{\mathrm v}$ by solving the
evolution equation. 

\subsubsection{Alternative choices of variables}

The perturbed conservation equations~\eqref{cons2.general} at second
order have been used to derive evolution equations
that differ from those in section~\ref{b_v.V_p}, but have the
unsatisfactory feature of not forming a closed system. First,
Fitzpatrick, Senatore, and Zaldarriaga (2010)~\cite{fitetal10}
have used the conservation equations in the Poisson gauge, first specialized to
the case of pressure-free matter and then to the case of radiation
(see equations (26)-(29)). The resulting
equations are first order evolution equations for ${}^{(2)}\!\bdelta_{\mathrm p}$
and ${}^{(2)}\!V_{\mathrm p}$, but are less
simple than equations~\eqref{delta_v,V_p_evol2} in that
they are coupled to the metric perturbation  ${}^{(2)}\!\psi_{\mathrm p}$.
As a second example Doran \emph{et al} (2003)~\cite{doretal03} apply
the perturbed conservation equations at linear order in a different way, using as variables
the density perturbation in the uniform curvature gauge ${}^{(1)}\!\bdelta_{\mathrm c}$
and the velocity perturbation in the Poisson gauge (see equations (A.29) and (A.30)).
Again these equations are coupled to the metric perturbation ${}^{(1)}\!\psi_{\mathrm p}$.

\section{Discussion\label{discussion}}

In this paper we have given five ready-to-use systems of governing equations
for second order scalar perturbations, subject to the assumption
that at first order the perturbations are purely scalar. Here we summarize their identifying
features and give their active dynamical variables:
\begin{itemize}
\item[i)] Equations~\eqref{p_gov2} using the Poisson gauge
(variable $\psi_{\mathrm p}$),
\item[ii)] Equations~\eqref{evol2_psi,V} using the Poisson gauge
(variables $\psi_{\mathrm p}, V_{\mathrm p}$),
\item[iii)] Equations~\eqref{ucg_gov2} using the uniform curvature gauge
(variables $\phi_\mathrm{c}, B_{\mathrm c}$),
\item[iv)] Equations~\eqref{totmat_2} using the total matter gauge
(variables $\psi_{\mathrm v}, B_{\mathrm v}$),
\item[v)] Equations~\eqref{delta_v,V_p_evol2} using
the conservation equations (variables $\bdelta_{\mathrm v}, {\bf D}^2 V_{\mathrm p}$).
\end{itemize}
Other systems of equations that are more general then ours, as regards matter content
and gauge choices, have been developed in the extensive series of papers
by Hwang and Noh (see for example~\cite{nohhwa04,hwanoh07a})
and Nakamura~\cite{nak07,nak10}. We regard our less general
but more focussed framework, which comprises
the above five ready-to-use systems of equations, as
complementing the more general systems in the above references.
Each of our systems is minimal in the sense
that there are no redundant equations or variables,
and the matter content is restricted
so that the systems are closed once the
non-adiabatic pressure perturbation $\Gamma$ is specified.
Although we are primarily motivated by the needs of second order
perturbation theory we note that our framework can be specialized
to linear perturbations by simply dropping the source terms. Because of this
we hope that our framework will
form a useful reference for both linear and second order perturbations.

We now make some remarks concerning the utility of the five systems of
governing equations as regards applications.
The unified nature of our formulation of these systems
of equations enables one to easily compare
their relative merits as regards
a chosen application. We begin by noting that
in cosmological perturbation theory the evolution of the
perturbations is described in general by partial
differential equations. Usually, in order to obtain explicit,
approximate or numerical solutions
in a particular physical context,
one applies the Fourier transform to the partial
differential equations
which converts them to ordinary differential equations for
the Fourier coefficients of the perturbation variables,
with the wave number $k$ as a parameter, together with
algebraic constraints relating the Fourier coefficients.
For first order perturbations
the spatial derivatives appear only via the spatial Laplacian ${\bf D}^2$,
and one can implement the transition by simply making the replacement
${\bf D}^2\rightarrow - k^2$. At second order, however,
the process is more complicated since one has to use
the Convolution Theorem to take
the Fourier transform of products of the
first order perturbations that appear in the source terms.\footnote{See
for example, Tram \emph{et al} (2016)~\cite{traetal16}, equations
(1.1)-(1.3) and Vretblad (2005)~\cite{vre05} for details.}

There are, however, two important applications of cosmological perturbation theory,
namely, adiabatic perturbations in the super-horizon regime
(the long wavelength limit) and perturbations of $\Lambda$CDM universes,
in which it is not necessary to make the transition to Fourier space
since the evolution equations  automatically simplify
to ordinary differential equations. We regard these applications
as elementary but important benchmark problems in cosmological perturbation theory.

First we note that long wavelength adiabatic perturbations
are defined by the requirement that terms of order 2 in
the scaled dimensionless spatial differential operator ${\cal H}^{-1}{\bf D}^i$
can be neglected, and that the non-adiabatic
pressure perturbation is negligible (${}^{(r)}\!\Gamma\approx 0,\,r=1,2$).
However, the background matter scalars $w$ and $c_s^2$ are unrestricted.
Second, as shown in appendix~\ref{frac.density.pert}, when the background model
is the $\Lambda CDM$ universe we have $w_m=0$ and hence
the background matter scalars are given by
\begin{equation}
c_s^2=0, \qquad  1+w=\Omega_m,
\end{equation}
which implies that the perturbations are adiabatic (${}^{(r)}\!\Gamma= 0,\,r=1,2$).
In both cases the term $c_s^2{\bf D}^2$, which appears in the
leading order terms in the evolution equations, is negligible,
and it is this property that reduces the evolution equations to
ordinary differential equations.

It turns out that in these two benchmark problems it is possible
to explicitly solve the ordinary differential equations and
obtain the general time dependence of the perturbations
at first and second order, including both growing and
decaying modes. The spatial dependence is described by arbitrary
spatial functions that arise as constants of integration. In order
to achieve this goal it is necessary to make an appropriate choice
from among the five ready-to-use systems. Observe that
in both systems iii) (uniform curvature gauge) and iv)
(total matter gauge) the two evolution equations decouple,
and can thus be solved successively for the two active variables,
first at linear order and then, after using the linear solution to
calculate the source terms, at second order. However, on evaluating
the source terms one finds that \emph{system iv), using
the total matter gauge, provides the
simplest method of solution for the two benchmark problems.}
Details concerning the derivation of the solution
in the case of long wavelength perturbations are given in
UW3~\cite{uggwai19b}.

We conclude with some brief remarks on the relative merits of the five systems for
problems other than the two benchmark problems. An immediate conclusion is that
system iv) is no longer the simplest system since the
presence of a non-zero $c_s^2$ complicates the evolution equations
considerably, since the term $c_s^2\bdelta_{\mathrm v}$,
which depends on $\psi_{\mathrm v} -{\cal H}B_{\mathrm v}$,
appears on the right side of both evolution equations
in~\eqref{totmat_1} and~\eqref{totmat_2}. In addition
the presence of this term makes the source terms more complicated.
Instead it appears that \emph{system iii), based on the uniform
curvature gauge, is the simplest system},
which makes it a natural choice for numerical experiments or
qualitative analysis using dynamical systems methods. This system
has not been given before.\footnote{We mention, however, that Malik and co-workers
have used the uniform curvature gauge to study
second order perturbations of inflationary
universes with single and multiple scalar fields (see for example, Malik (2007)~\cite{mal07},
Huston and Malik (2009)~\cite{husmal09}
and Christopherson \emph{et al} (2015)~\cite{chretal15}.)
The structure of the governing equations in these references is
specifically adapted to the scalar fields and as a result they do
not have much in common with our governing equations.} 					
Our analysis suggests that it is worthy of further study.

\begin{appendix}

\section{The general source terms\label{source}}

The major technical problem in second order perturbation theory is
managing the quadratic source terms. Our strategy is to use a consistent
and easy to identify notation.
We use the same letter for the kernel in the symbol  for the
source terms as we do for the leading order terms but with a different
font: ${\sf G}$ and ${\mathbb G}$ for the leading order and source terms
of the Einstein tensor and ${\sf T}$ and ${\mathbb T}$ for the
stress-energy tensor with superscripts $\Gamma,\pi,q,\rho$
indicating the components and
subscripts ${\mathrm p},{\mathrm c},{\mathrm v} $ indicating the gauge,
as in section~\ref{pert_Einst}.  For the conserved energy and momentum
equations we use ${\sf E}$ and ${\mathbb E}$ and
${\sf M}$ and ${\mathbb M}$, respectively, with the usual subscripts
indicating the gauge. We also introduce a notation for the source terms
associated with the change of gauge formulas at second order, ${\mathbb S}(\dots)$,
where $(\dots)$ identifies the formula (see equations~\eqref{B_v+V_p}
and~\eqref{delta_p-3HV_p-delta_v}). Other source terms  are
defined as linear combinations of the above basic expressions (see
sections~\ref{b_v.V_p} and~\ref{evol.delta_v}).

In writing the source terms it is convenient to follow appendix B in
UW1~\cite{uggwai19a}
and define the following spatial differential operators:
\begin{subequations}
\begin{align}
({\bf D} C)^2 &= \gamma^{ij}{\bf D}_iC{\bf D}_jC, \label{QD1}  \\
{\mathbb D}_0(C) &= {\cal S}^{ij}{\bf D}_iC{\bf D}_jC, \label{QD2} \\
{\mathbb D}_2(C) &= \sfrac13\left({\bf D}^2{\mathbb D}_0(C)- ({\bf D}C)^2\right),\label{QD3}
\end{align}
\end{subequations}
where the scalar mode extraction operators ${\cal S}^i$ and ${\cal S}^{ij}$
are given by~\eqref{modeextractop}.

By inspection of the expressions for the source terms one finds that
$\cal H$ appears explicitly only in the variables ${\cal H}B$ and ${\cal H}V$
 and as a coefficient of the spatial differential operator, in the
 form ${\cal H}^{-1}{\bf D}_i$.
We can thus absorb all multiplicative factors of  ${\cal H}$ by introducing
the following overbar notation:
\begin{equation}  \label{bar_notation1}
\bar{B} = {\cal H}B, \qquad \bar{V} = {\cal H}V, \qquad
\bar{\bf D}_i = {\cal H}^{-1}{\bf D}_i,
\end{equation}
thereby making the expressions for the source terms simpler.
The definition of $\bar{\bf D}_i$ leads to barred expressions for the
associated spatial differential operators:
\begin{subequations}   \label{bar_notation2}
\begin{alignat}{2}
\bar{\bf D}^2 &= {\cal H}^{-2}{\bf D}^2, &\qquad
\bar{\bf D}_{ij} &= {\cal H}^{-2}{\bf D}_{ij},  \\
(\bar{\bf D} C)^2 &= {\cal H}^{-2}({\bf D} C)^2, &\qquad
\bar{\mathbb D}_2(C) &= {\cal H}^{-2}{\mathbb D}_2(C),  \\
\bar{\cal S}^i &= {\cal H}{\cal S}^i, &\qquad
\bar{\cal S}^{ij} &= {\cal H}^{2}{\cal S}^{ij}.
\end{alignat}
\end{subequations}
We could also use the barred dimensionless expressions to simplify the
terms in the ready-to-use systems of governing equation in section~\ref{gov.eq}
but have decided to give the more familiar forms in which ${\cal H}$ is visible.


\subsection{The Einstein tensor source terms}

The expressions for the Einstein source terms
in~\eqref{einst_pert2_lincombo} can be obtained
by specializing equations (75) -(78) in Uggla and
Wainwright (2013)~\cite{uggwai13a}.\footnote{See
footnote~\ref{link.2013} for the relation between the notation
in~\cite{uggwai13a} and in the present paper.}
 With the above notation the source terms can be written in the following
form:
\begin{subequations}\label{source_Gab}
\begin{align}
{\mathbb G}^{\rho}({}^{(1)}\!f) &= 6[4\phi^2 -
(\bar{\bf D}\bar{B})^2] + {\mathbb W} - 2{\mathbb X} +
2\bar{\bf D}^2\bar{\mathbb D}_2(\bar{B}), \label{source_Gab.1}   \\
\begin{split}
{\mathbb G}^{\Gamma}({}^{(1)}\!f) &= -2{\cal L}_1[4\phi^2 - (\bar{\bf D}\bar{B})^2] -
\sfrac13(2 + 3{\cal C}^2){\mathbb W} + \sfrac23(1 + 3{\cal C}^2){\mathbb X}\\
& \quad - \sfrac23{\mathbb R} - \sfrac23(1 + 3{\cal C}^2)\bar{\bf D}^2\bar{\mathbb D}_2(\bar{B}),
\end{split}\\
{\mathbb G}^q({}^{(1)}\!f) &= 2[4\phi^2 - (\bar{\bf D}\bar{B})^2] +
8(\phi - \psi)\partial_N\psi + \bar{\cal S}^i {\mathbb R}_i, \label{source_Gab.3}  \\
\begin{split}
{\mathbb G}^{\pi}({}^{(1)}\!f) &= 4\psi^2 + 2{\mathbb D}_0(\phi) - 6{\mathbb D}_0(\psi)  \\
&\quad  + 2\bar{\cal S}^{ij}[2(\phi-\psi)\bar{\bf D}_{ij}\psi +2(\bar{\bf D}_i \phi)\bar{\bf D}_{j}\bar{B} +
\left(\partial_N(\phi + \psi)\right)\bar{\bf D}_{ij}\bar{B}]  \\
&\quad + 4\bar{\cal S}^{ij}[(\psi - \phi)\bar{\bf D}_{ij}{\mathsf G}^{\pi}({}^{(1)}\!f) +
(\bar{\bf D}_i\psi)\bar{\bf D}_{j}{\mathsf G}^{\pi}({}^{(1)}\!f)]  + \bar{\mathbb D}_2(\bar{B}), \label{source_Gab.4}
\end{split}
\end{align}
\end{subequations}
%
%
%
where
\begin{subequations}\label{source_Gab_extra}
\begin{align}
{\mathbb W} &= 24(\phi - \psi)\partial_N\psi + 8(\phi - \psi)\bar{\bf D}^2\bar{B} +
4(\bar{\bf D}_i\bar{B})\bar{\bf D}^i(\phi + \psi),\\
{\mathbb X} &=  - 3(\partial_N\psi)^2 + 5(\bar{\bf D}\psi)^2 - 4\bar{\bf D}^2\psi^2 -
2\bar{\bf D}^i\!\left((\bar{\bf D}_i\bar{B})\partial_N\psi\right),\\
\begin{split}
{\mathbb R} &= 12(\phi - \psi)(\partial_N^2 - q\partial_N)\psi
+ 6(\partial_N\psi)\partial_N(\phi - \psi) + 4(\phi - \psi)\bar{\bf D}^2\psi \\
& \quad + 2(\bar{\bf D}\phi)^2 + 2(\bar{\bf D}\psi)^2 +
4(\bar{\bf D}_i\bar{B})\bar{\bf D}^i\partial_N\psi +
2(\partial_N\phi){\bf D}^2\bar{B}\\
& \quad - 2[2(\phi - \psi)\bar{\bf D}^2 + (\bar{\bf D}_i\psi)\bar{\bf D}^i]
[2\bar{B} + {\mathsf G}^{\pi}(f)],
\end{split}\\
\begin{split}
{\mathbb R}_i &=  -4(\partial_N\psi)\bar{\bf D}_i \phi +
2(\bar{\bf D}_j\bar{B})(\bar{\bf D}^j\!_i + \sfrac43\delta^j\!_i\bar{\bf D}^2)\psi -
2(\bar{\bf D}_j\phi)(\bar{\bf D}^j\!_i - \sfrac23\delta^j\!_i\bar{\bf D}^2)\bar{B} .
\end{split}
\end{align}
\end{subequations}
%
%
%
Here and elsewhere in the Appendix, in order to simplify the notation
 we have omitted the superscript ${}^{(1)}$ on the linear perturbations
in the source terms.
We have simplified the term~\eqref{source_Gab.4} in an  important  way, as follows.
This term initially contains the expression ${\mathbb D}_2^{*}(B)$, where
\begin{equation}
{\mathbb D}_2^{*}(B) :=
2{\cal S}^{ij}\left(\sfrac13({\bf D}^2 B){\bf D}_{ij}B - ({\bf D}_{k\la i} B)\,{\bf D}^k\!_{j\ra}B\right),
\end{equation}
but in the flat case it can be shown using the commutation identities for ${\bf D}_i$
that
\begin{equation}
{\mathbb D}_2^*(B) =  {\mathbb D}_2(B).
\end{equation}
%

\subsection{The stress-energy tensor source terms}

The components of the source term, identified by a kernel ${\mathbb T}$, are given by
\begin{subequations}  \label{source_Tab}
\begin{align}
{\mathbb T}^{\rho} &= \gamma^{ij}({\mathbb V}_2)_{ij},  \label{source_Tab.1}   \\
{\mathbb T}^{\Gamma} &= \sfrac13(1-3c_s^2)\gamma^{ij}({\mathbb V}_2)_{ij}
- \sfrac13(\partial_N c_s^2)\bdelta^2 -
\sfrac23 {\bdelta}\left(\partial_N - 3(1+c_s^2)\right)\!{\Gamma}, \\
{\mathbb T}^q &=
\bar{\cal S}^i\!\left[2\left((1+c_s^2)\bdelta + \Gamma - \phi\right)\bar{\bf D}_i\bar{V}\right],
\label{source_Tab.3} \\
{\mathbb T}^{\pi} &= \bar{\cal S}^{ij}({\mathbb V}_2)_{ij},
\end{align}
where
\begin{equation}
({\mathbb V}_2)_{ij} :=  2({\bf D}_i{V}){\bf D}_j \left({V} - {B}\right) =
2(\bar{\bf D}_i\bar{V})\bar{\bf D}_j \left(\bar{V} - \bar{B}\right),
\end{equation}
\end{subequations}
where $({\mathbb V}_2)_{ij}$ has weight $2$ in ${\bf D}_i$,
motivating the subscript ${}_2$.

\subsection{The source terms for the conservation equations}

The quadratic source terms are given by
\begin{subequations} \label{source_divT}
\begin{align}
\begin{split}
{\mathbb E}({}^{(1)}\!F) &= -\partial_N[6\psi^2 -
(\bar{\bf D}\bar{V})^2 + (1 + c_s^2)\bdelta^2 + 2\bdelta\Gamma]  - 6\Gamma^2\\
& \quad + 2\bar{\bf D}^k[\phi\bar{\bf D}_k\bar{V} + 2\psi\bar{\bf D}_k(\bar{V} - \bar{B})] +
2[\bar{\bf D}^k(\bdelta - 3\psi)]\bar{\bf D}_k(\bar{V} - \bar{B}),
\end{split}   \label{E_source_divT}   \\
\begin{split}
{\mathbb M}({}^{(1)}\!F) &= -2\phi^2 + [\bar{\bf D}(\bar{V} - \bar{B})]^2 -
 \left[c_s^2(1 + c_s^2) + \sfrac13(\partial_N c_s^2)\right]\!\bdelta^2
- \Gamma^2 - \sfrac23\bdelta\partial_N\Gamma \\
& \quad -2\bar{\cal S}^i\!\left\{\left[\phi(\partial_N + 1 - 3c_s^2) -
\partial_N(c_s^2\bdelta + \Gamma) + 3\Gamma \right]\!\bar{\bf D}_i\bar{V}
- \Gamma\bar{\bf D}_i\bdelta\right\},
\end{split}   \label{M_source}
\end{align}
\end{subequations}
where we have used the first order equation $\mathsf{E}({}^{(1)}\!F) = 0$
in deriving the first equation.

The source term ${\mathbb S}_{Poisson}$ in the
GR version of Poisson's equation
at second order~\eqref{M.2}, that is used in the source term~\eqref{source_V},
has the following form:
\begin{equation} \label{S_Poisson}
{\mathbb S}_{Poisson}={\bf \bar D}^2\left(4\psi_{\mathrm p}^2
-(1+q) (\bar V_{\mathrm p})^2 \right)-
5({\bf\bar  D}\psi_{\mathrm p})^2 -
 6{\cal \bar S}^i \left[\bar V_{\mathrm p}{\bf \bar D}_i ({\bf \bar D}^2\psi_{\mathrm p})\right].   \end{equation}
%

\subsection{The source terms for the change of gauge formulas}

We have recently given in UW1~\cite{uggwai19a} a general
formalism for relating gauge invariants
associated with different gauges at second order. In this paper
we need the gauge change formula that relates $\bar{B}_{\mathrm v}$
to $\bar{V}_{\mathrm p}$ which we write in the following form
using the bar notation:
\begin{equation} \label{B_v+V_p}
{}^{(2)}\!\bar{B}_{\mathrm v} + {}^{(2)}\!\bar{V}_{\mathrm p}=
{\mathbb S}(\bar{B}_{\mathrm v} + \bar{V}_{\mathrm p}),
\end{equation}
and also the one that relates ${}^{(2)}\!{\bdelta}_{\mathrm v}$
to ${}^{(2)}\!{\bdelta}_{\mathrm p}$:
\begin{equation} \label{delta_p-3HV_p-delta_v}
{}^{(2)}\!\bdelta_{\mathrm p} - 3{}^{(2)}\!\bar{V}_{\mathrm p} -
{}^{(2)}\!{\bdelta}_{\mathrm v} = {\mathbb S}(\bdelta_{\mathrm p} -
3\bar{V}_{\mathrm p} - {\bdelta}_{\mathrm v}).
\end{equation}
The source terms are given by
\begin{subequations}  \label{gauge.change.source}
\begin{align}
{\mathbb S}(\bar{B}_{\mathrm v} + \bar{V}_{\mathrm p}) &=
(\partial_N + 2q)[{\mathbb D}_0(\bar{V}_{\mathrm p}) - \bar{V}_{\mathrm p}^2] -
2\bar{V}_{\mathrm p}^2 -
2\bar{\cal S}^i[\phi_{\mathrm p}\bar{\bf D}_i\bar{V}_{\mathrm p}], \label{B+V.source}\\
\begin{split}
{\mathbb S}(\bdelta_{\mathrm p} -
3\bar{V}_{\mathrm p} - {\bdelta}_{\mathrm v}) &=
3(1 + q + 3(1+c_s^2))\bar{V}_{\mathrm p}^2 -
6\bar{\cal S}^i[\phi_{\mathrm v}\bar{\bf D}_i\bar{V}_{\mathrm p}] \\ & \qquad
+ 2(3{\bdelta}_{\mathrm v} +
 \bar{\bf D}^2\bar{V}_{\mathrm p})\bar{V}_{\mathrm p} .
  \label{delta_p,v_source}
\end{split}
\end{align}
\end{subequations}
For the first equation we used equation (42a) in~\cite{uggwai19a} with
$\Box = \bar{V} = {\cal H}V$ and the total matter gauge on the right side,
and for the second equation we used equation (42b) with
$\Box=\sfrac13 {\bdelta}$ and the Poisson gauge on the right side.
It is then necessary to use the definitions (36) of the hatted variables.

\section{The fractional density perturbation \label{frac.density.pert}}

We emphasize that $\bdelta$ is defined by normalizing the density perturbation
with $\rho_0+p_0$ as in equation~\eqref{rho,p_normalized}, while
the commonly used fractional density perturbation $\delta$ is defined
by normalizing the density perturbation with the background matter density,
which we denote by ${}^{(0)}\!\rho_{m}$, while $\rho_0$ denotes the total matter/energy
density. If there is a cosmological constant then
\begin{equation}
{}^{(0)}\!\rho = \rho_0= {}^{(0)}\!\rho_{m} + \Lambda, \qquad {}^{(0)}\!p = p_0= {}^{(0)}\!p_{m} - \Lambda,
\end{equation}
and we also introduce
\begin{equation} \label{w_m}
w_m= \frac{{}^{(0)}\!p_{m}}{{}^{(0)}\!\rho_{m}}, \qquad \Omega_m = \frac{{}^{(0)}\!\rho_{m}}{3H^2},
\end{equation}
while $c_s^2$ is unaffected. It follows that
\begin{equation} \label{w.w_m}
1+w = \Omega_m(1+w_m).
\end{equation}
The fractional density perturbation is defined by
\begin{equation}
{}^{(r)}\!\delta = \frac{{}^{(r)}\!\rho}{{}^{(0)}\!\rho_{m}}, \quad r=1,2.
\end{equation}
It follows that $\delta=(1+w_m)\bdelta$, which simplifies
to $\delta=(1+w)\bdelta$ if $\Lambda=0.$
In particular for a $\Lambda CDM$ universe we have $w_m=0,\, c_s^2=0$
and~\eqref{w.w_m} reduces to
\begin{equation} \label{LCDM}
1+w=\Omega_m.
\end{equation}

\end{appendix}

\bibliographystyle{plain}
\bibliography{../Bibtex/cos_pert_papers}

\end{document}